\documentclass[fleqn,usenatbib]{mnras}
\usepackage[T1]{fontenc}
\usepackage{ae,aecompl}
\usepackage{graphicx}	
\usepackage{subfig}
\usepackage{epstopdf}
\usepackage{amsmath}	
\usepackage{amssymb}	
\usepackage{xspace} 
\usepackage{natbib}
\usepackage[bottom]{footmisc}
\usepackage{hyperref}

\newcommand{\spitzer}{\textit{Spitzer}\xspace}
\newcommand{\msun}{\hbox{$\hbox{\rm M}_{\odot}$}\xspace}
\newcommand{\herschel}{\textit{Herschel}\xspace}
\newcommand{\lsun}{L$_{\odot}$\xspace}

\newcommand{\kms}{\rm km\,s^{-1}}
\newcommand{\pc}{{\rm pc}\xspace}
\newcommand{\yr}{{\rm yr}\xspace}
\newcommand{\myr}{{\rm Myr}\xspace}

\newcommand{\mum}{\micron\xspace}
\newcommand{\degree}{$^{\circ}$}
\newcommand{\ammonia}{${\rm NH}_3$\xspace}
\newcommand{\nthp}{${\rm N}_2{\rm H}^+$\xspace}

\title[Velocity structure of the ISF]{Gas velocity structure of the Orion A Integral Shaped Filament}

\author[Gonz\'alez \& Stutz]{
Valentina Gonz\'alez Lobos,$^{1}$\thanks{E-mail: vgonzalezl@udec.cl}
Amelia M.\ Stutz$^{1,2}$
\\
$^{1}$Departmento de Astronom\'{i}a, Facultad de Ciencias F\'{i}sicas y Matem\'{a}ticas, Universidad de Concepci\'{o}n, Concepci\'{o}n, Chile\\
$^{2}$Max-Planck-Institute for Astronomy, K\"onigstuhl 17, 69117 Heidelberg, Germany}

\date{Accepted 2019 August 30. Received 2019 August 30; in original form 2019 June 20}

\pubyear{2019}

\begin{document}
\label{firstpage}
\pagerange{\pageref{firstpage}--\pageref{lastpage}}
\maketitle

\begin{abstract}
We present analysis of the gas kinematics of the Integral Shaped Filament (ISF) in Orion~A using four different molecular lines, $^{12}$CO (1-0), $^{13}$CO (1-0), \ammonia(1,1), and \nthp(1-0).  We describe our method to visualize the position-velocity (PV) structure using the intensity-weighted line velocity centroid, which enables us to identify structures that were previously muddled or invisible.  We observe a north to south velocity gradient in all tracers that terminates in a velocity peak near the center of the Orion Nebula Cluster (ONC), consistent with the previously reported ``wave-like'' properties of the ISF.  We extract the velocity dispersion profiles and compare the non-thermal line widths to the gas gravitational potential.  We find supersonic Mach number profiles, yet the line widths are consistent with the gas being deeply gravitationally bound.  We report the presence of two $^{12}$CO velocity components along the northern half of the ISF; if interpreted as circular rotation, the angular velocity is $\omega=1.4\,{\rm Myr}^{-1}$.  On small scales we report the detection of \nthp and \ammonia ``twisting and turning'' structures, with short associated timescales that give the impression of a torsional wave.  Neither the nature of these structures nor their relation to the larger scale wave is presently understood.  
\end{abstract}


\section{Introduction}

The Orion~A molecular cloud is the largest nearby \citep[$d\sim400\,\pc$;  e.g.,][]{schlafly14,stutz2019,kounkel2018,gross2018,zari18,getman2019} molecular cloud hosting massive star and cluster formation.  The proximity of the Orion cloud allows for detailed scrutiny of its stellar content \citep[e.g.,][]{kroupa00,hillenbrand98,tobin09,port2016,kounkel2018}, young stellar object (YSO) and protostellar content \citep[e.g.,][]{megeath2012,stutz13,furlan2016,kainulainen2017,stutz2016,stutz2018a,stutz2019}, and its gas content \citep[e.g.,][]{tatematsu08, nishimura15, stutz2015,stutz2016,stutz2019,hacar17,friesen2017,kong18}.  Meanwhile, wide-field scrutiny of the magnetic fields in the gas \citep[e.g.,][]{heiles97,matthews00} has recently received renewed attention \citep[e.g.,][]{pattle17,tahani18,reissl18,dom18,chuss19} driven by improvements in observing and simulation capabilities.  
Generally the above studies have, when approaching the subject of the dynamics of the stars and/or gas, proceeded without knowledge of the total gas mass and gas potential, which has only been made available recently \citep[][]{stutz2015,stutz2016,stutz2018a}.  

The Orion~A molecular cloud can be divided into two principal regions: the Integral Shaped Filament \citep[ISF; ][]{bally87,stutz2016} in the north, and the L1641 cloud in the south.  The Orion Nebula Cluster (ONC; see references above) is forming out of the ISF.  The ISF is thus distinct from the southern more diffuse L1641 region in a few key characteristics: the ISF is host to massive star and cluster formation, has a higher gas mass per unit length, is denser, has a different gravitational potential profile, and likely has different magnetic field properties \citep[][]{stutz2016,stutz2018a} compared to L1641. 

Based on the comparison between stellar radial velocities (measured from the APOGEE survey) and the gas potential, \citet{stutz2016} proposed the Slingshot model, in which the ISF gas is undergoing oscillations driven by the interaction between the magnetic field and gravity.  In the Slingshot, the stars, which are born on the filament ridgeline, are ejected from their cradles as the filament oscillates.  Subsequently, \citet{stutz2018a} proposed a similar Slingshot scenario for the ONC as a whole. In these works, the gas and stellar mass distributions were measured from the data, while the gravitational potential was estimated using simple geometric assumptions.  That is, cylindrical and spherical geometries were assumed for the gas and stars, respectively, in order to deproject the observed quantities.

A key piece of evidence in the Slingshot scenario was the observed radial velocity gradient in the ISF gas, which was quantified by \citet{stutz2016} thanks to the \citet[][]{nishimura15} CO maps and the higher density \nthp maps from \citet[][]{tatematsu08}.  Subsequently, in independent work, \citet[][]{kong18} presented a higher  resolution version of the ISF CO gas distribution based on interferometer data; these authors proposed that the  morphology of their position-velocity diagrams was indeed consistent with a wave-like perturbation in the gas, as proposed in the Slingshot.  Thus, the kinematic structure and properties of the gas, and specifically possible wave-like nature of the ISF, is a key test of the Slingshot.  In order to address this, \citet[][]{stutz2019} compared the Gaia parallaxes of young stars to the \nthp gas velocities and found that the ISF has properties to first order consistent with a standing wave, where the YSO displacement is proportional to the gas radial velocity.  Recently, the Gaia measurements in \citet[][]{stutz2019} were confirmed by \citet{getman2019}.  

In this work we focus exclusively on the gas kinematics of the ISF and present detailed analysis based on  public data of four molecular lines, $^{12}$CO~(1-0), $^{13}$CO~(1-0), \ammonia~(1,1), and \nthp~(1-0).  We show that the ISF has complex gas structure, including indications of possible large scale rotation and small-scale twistings and turnings reminiscent of the action of a torsional wave.  We introduce intensity-weighted position-velocity diagrams as a way to highlight existing kinematic structures and features in the gas data.  We compare the gas line widths to previously derived gas gravitational potentail of the ISF and the ONC and show that the gas linewidths are consistent with the cloud being gravitationally bound.  This paper is organized as follows. In \S~\ref{sec:data} we describe the observations; in \S~\ref{sec:pv} we describe the data analysis and the intensity-weighted position-velocity diagrams; in \S~\ref{sec:results} we present our main results; and in \S~\ref{sec:sum} we summarize our results. 

\section{Data}\label{sec:data}

Below we describe each data-set that we use in this work.  All data are publicly available and focus on the Integral Shaped Filament \citep[ISF; e.g.,][]{bally87, stutz2016} region of the Orion~A molecular cloud, where the Orion Nebula Cluster (ONC), also known as M42, is forming.

\subsection{Gas and stellar mass maps}

To quantify the gas mass distribution, we use the column density N(H) map from \citet{stutz2018a}. This map was derived from the \herschel dust emission at 160~\mum, 250~\mum, 350~\mum, and 500~\mum \citep{stutz2015} with a resolution of $\sim\,$20\,\arcsec, corrected in the center of the ONC using APEX 870~\mum data \citep[e.g.,][]{stanke10,stutz13} due to saturation \citep{stutz2018a}. The N(H) data correspond to the total hydrogen column density, where $N({\rm H}) = 2\times N(\text{H}_2)$ (see \citealt{stutz2015} for more details).  For the stellar mass distribution we use the \citet{stutz2018a} stellar mass map.  This map was constructed by counting \spitzer-identified young stellar objects (YSOs), from \citet{megeath2016}. To compute the stellar mass, \citet{stutz2018a} assumes a disk fraction of 0.75 \citep{megeath2016} and a mean stellar mass of $0.5\,\msun$ \citep{kroupa01}.

\subsection{ $^{12}$CO (1-0) and $^{13}$CO (1-0)  data}

\begin{figure}
\centering
\includegraphics[width=0.45\textwidth]{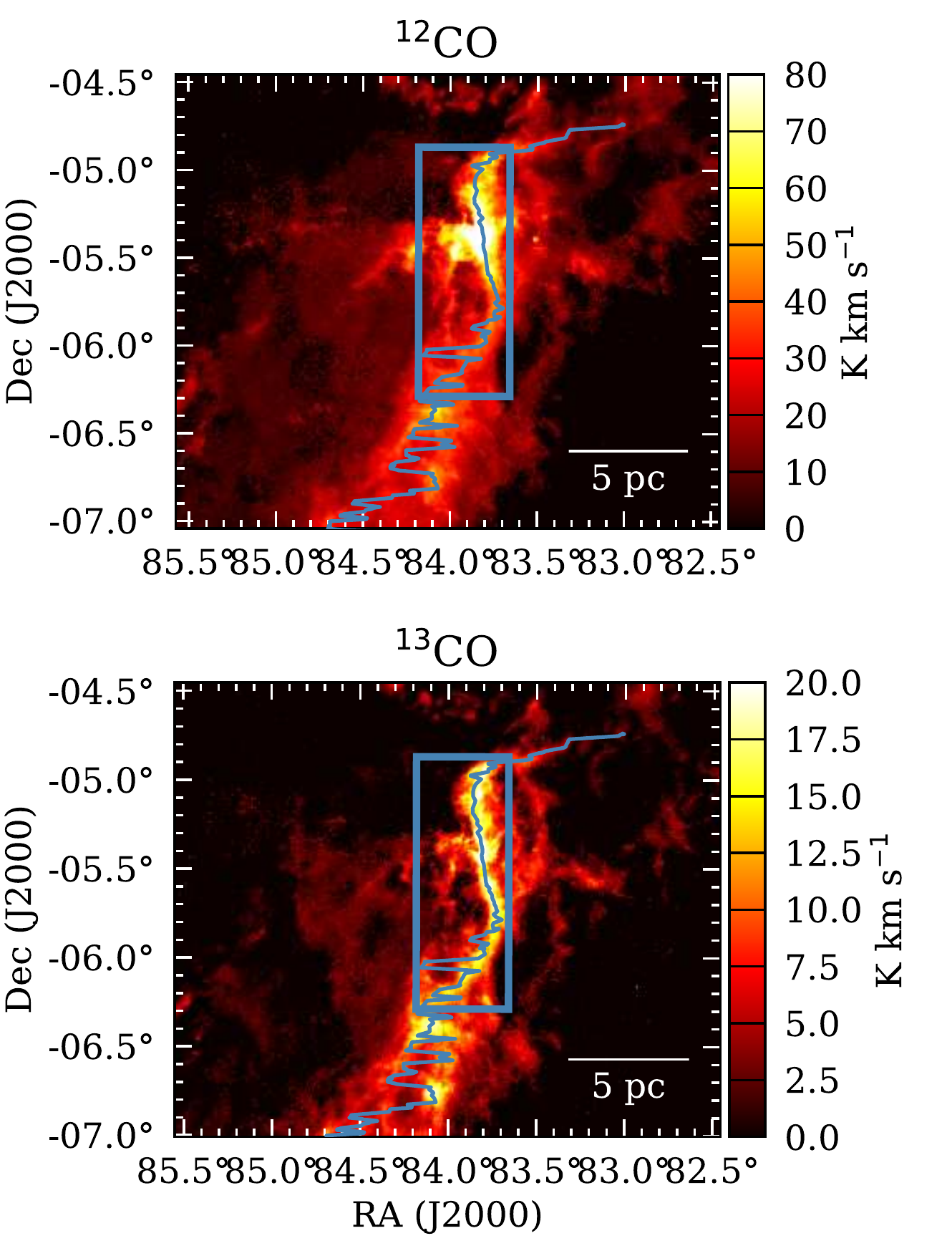}
\caption{$^{12}$CO (1-0) (top) and $^{13}$CO (1-0) (bottom) integrated intensity maps of the Orion~A region, from \citet{ripple2013}. The maps were integrated over a velocity range from ${\rm v}_{\rm LSR}=0\,\kms$ to ${\rm v}_{\rm LSR}=16\,\kms$. In both panels the blue curve is the column density ridgeline \citep{stutz2018a} and the blue box indicates the region that the \nthp observation field covers, that is, the Integral Shaped Filament (ISF). A $5\,\pc$  scale-bar is shown at the bottom of each panel.}
\label{fig:co_maps}
\end{figure}

We use the $^{12}$CO (1-0) and $^{13}$CO (1-0) observations from \citet{ripple2013}. The observations were carried out with the Five College Radio Astronomy Observatory 14 meter telescope during 2005 and 2006. The Full Width Half Maximum (FWHM) of the beam is 45$^{\prime\prime}$ for $^{12}$CO and 47$^{\prime\prime}$ for $^{13}$CO. The spectral resolution of each data cube is ${0.077\,\kms}$  for $^{12}$CO and ${0.08\,\kms}$ for $^{13}$CO. The $1\sigma$ errors due to thermal noise within $0.2\,\kms$ wide channels are 2.0 K for $^{12}$CO and 0.77 K for $^{13}$CO, as reported in \citet{ripple2013}. We refer the reader to \citet{ripple2013} for further details of this dataset. The critical density of the $^{12}$CO (1-0) transition is about 740 cm$^{-3}$ and 1.9$\times 10^{3}$ cm$^{-3}$ for $^{13}$CO (1-0) \citep{wilson2009, hernandez2011}. Figure \ref{fig:co_maps} shows the velocity integrated intensity maps of $^{12}$CO and $^{13}$CO respectively.

\begin{figure}
\centering
\includegraphics[width=0.47\textwidth]{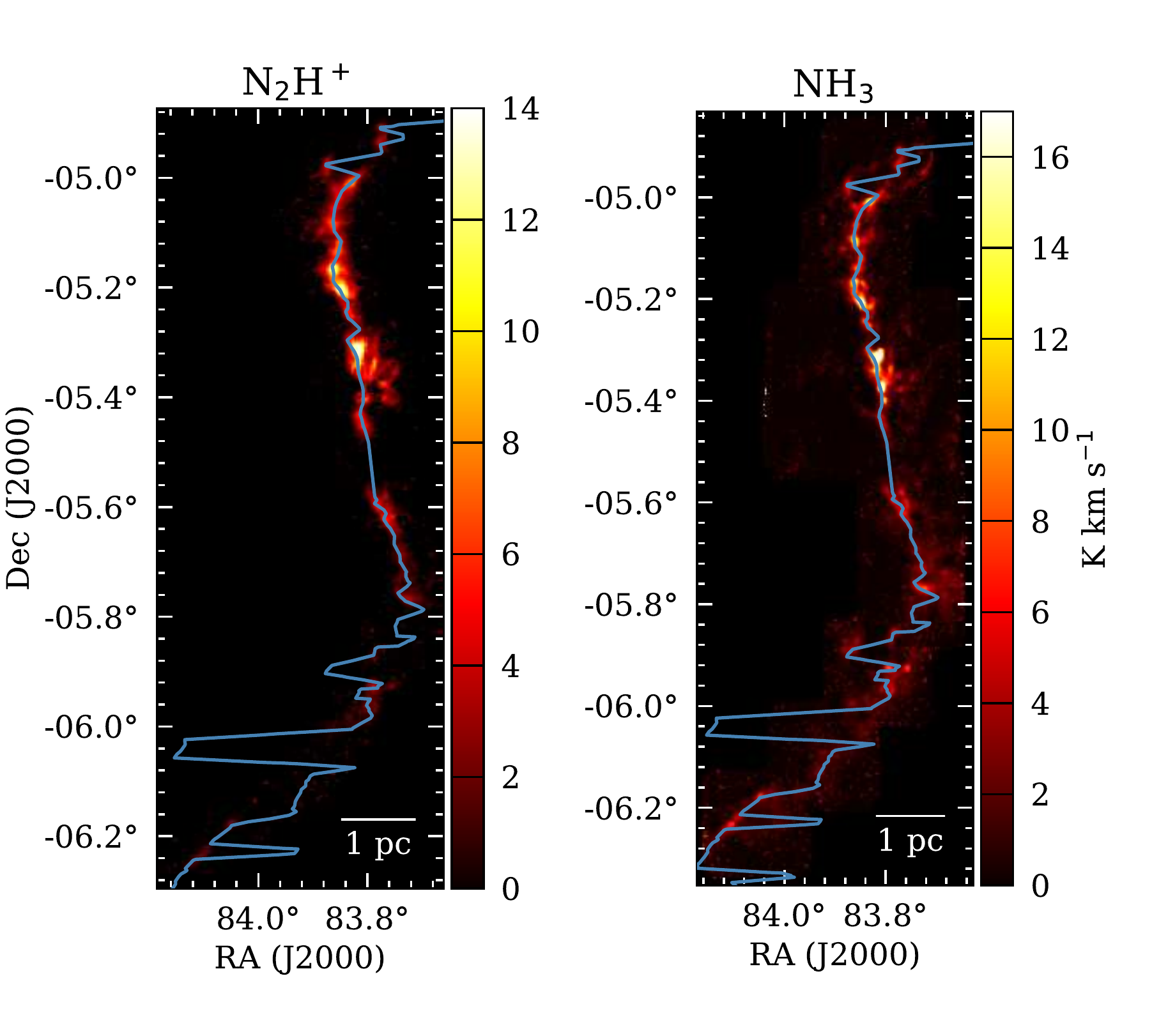}
\caption{\textit{Left}: \nthp (1-0) integrated intensity map of the ISF region from \citet{tatematsu08}. \textit{Right}: \ammonia (1,1) velocity integrated emission map of the ISF region from \citet{friesen2017}. A $1\,\pc$ scalebar is shown at the bottom of each panel.  The blue curve is the column density ridgeline \citep{stutz2018a}.  The area covered by the \nthp map corresponds to the blue box in Figure~\ref{fig:co_maps}.}
\label{fig:n2hp_nh3}
\end{figure}

\subsection{\texorpdfstring{\nthp}%
     {TEXT} (1-0) data}

 We use observations of \nthp (1-0) of \cite{tatematsu08} carried out with the Nobeyama Radio Observatory 45\,m telescope between 2005 and 2007. The FWHM is $\sim17^{\prime\prime}$ at 93\,GHz. The spectral resolution of this data cube is ${0.12\,\kms}$. The \nthp transition has a critical density of 6.1${\times 10^4}$\,cm$^{-3}$ assuming a kinetic temperature of 10\,K \citep{shirley2015}.  We refer the reader to \citet{tatematsu08,tatematsu2016} for further details. The left pannel of Figure \ref{fig:n2hp_nh3} shows the velocity integrated emission of \nthp, which covers the ISF in the northern portion of the Orion~A cloud.  

\subsection{\texorpdfstring{\ammonia}%
    {TEXT} data}

We use the first public release of the Green Bank Ammonia Survey (GAS) observations \citep{friesen2017} of the Orion~A North region. These correspond to the \ammonia~(1,1), (2,2) and (3,3) transitions with a spectral resolution of ${\sim0.07\,\kms}$, a typical $1\sigma$ noise value of $<0.1K$ and beamsize of $\sim32^{\prime\prime}$ at 23.7\,GHz, corresponding to the \ammonia~(2,2) rest frequency. For more details on the observations and analysis see \citet{friesen2017}. \ammonia has a critical density of $\sim 2\times10^3\,{\rm cm}^{-3}$ assuming a kinetic temperature of 10\,K \citep{shirley2015}. The resulting \citet{friesen2017} maps of interest here are the integrated intensity and centroid velocity maps.  These maps were obtained from \ammonia line fitting analysis using the Python package PySpecKit \citep{pyspeckit}. They perform the fit on \ammonia (1,1) lines where the signal to noise ratio is $>3$. We refer the reader to \S~3 of \citet{friesen2017} for further details in the \ammonia line fitting. The right pannel of Figure~\ref{fig:n2hp_nh3} shows the \ammonia~(1,1) integrated intensity in the ISF \citep{friesen2017} .

\subsection{Point sources}\label{sec:pro_cat}

We use several protostar, YSO, and point source catalogs to compare their positions to the radial velocity structure of the ISF (see \S~\ref{sec:velocity_struct}). The protostar selection is limited to the ISF region: the Herschel Orion Protostar Survey (HOPS, \citealt{stutz13,furlan2016}), protostellar cores observed with ALMA at $3\,{\rm mm}$ continuum of the northern part of the ISF from \cite{kainulainen2017}, protostellar cores identified in the OMC1 North region observed with the Submillimeter Array (SMA) at 1.3\,mm from \cite{teixeira2016}, and the outflow source in the Orion KL region from \cite{zapata2009} observed with the SMA.

\begin{figure*}
\centering
\subfloat{\centering\includegraphics[width=0.48\textwidth]{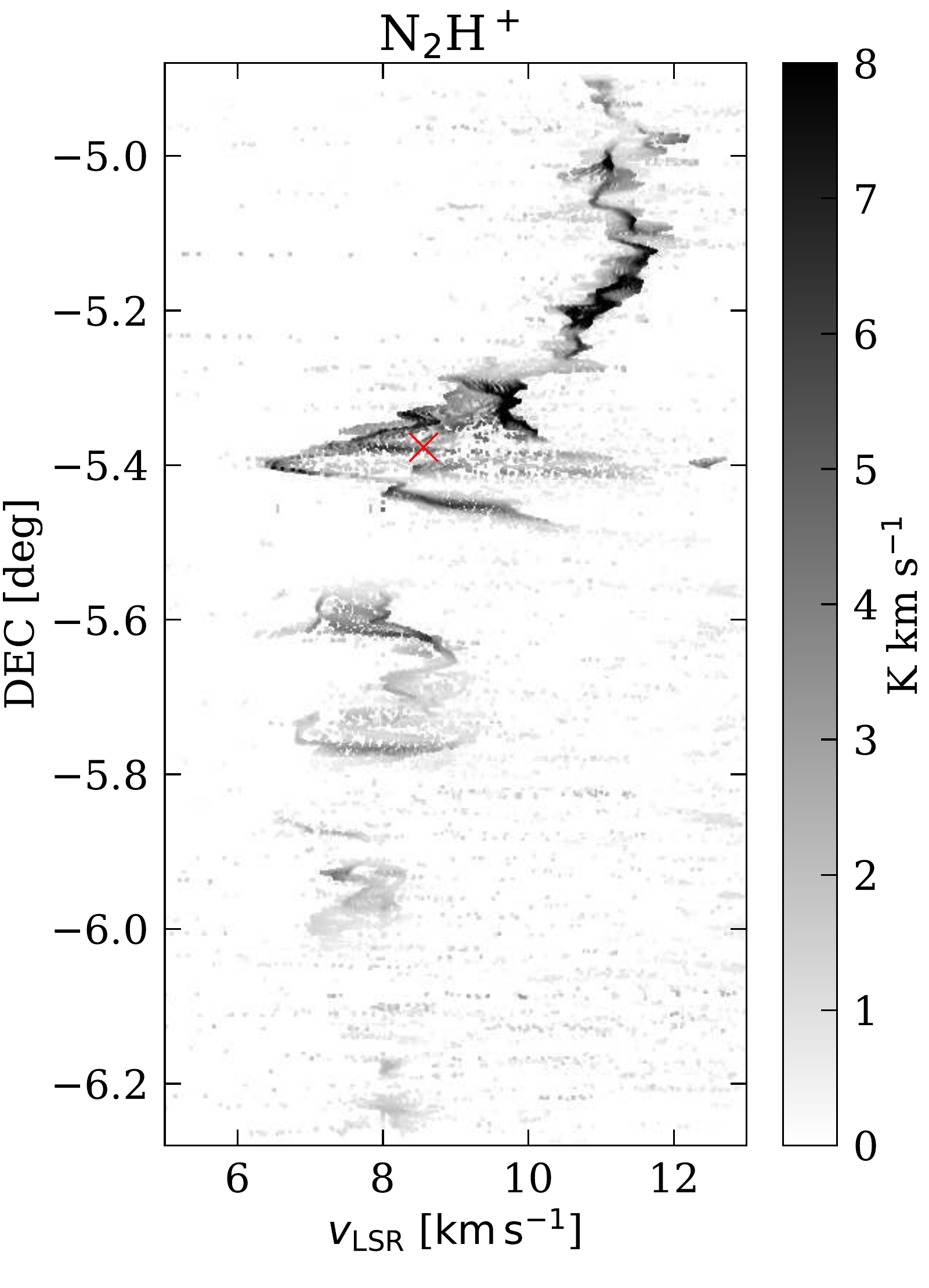}}
\subfloat{\centering\includegraphics[width=0.41\textwidth]{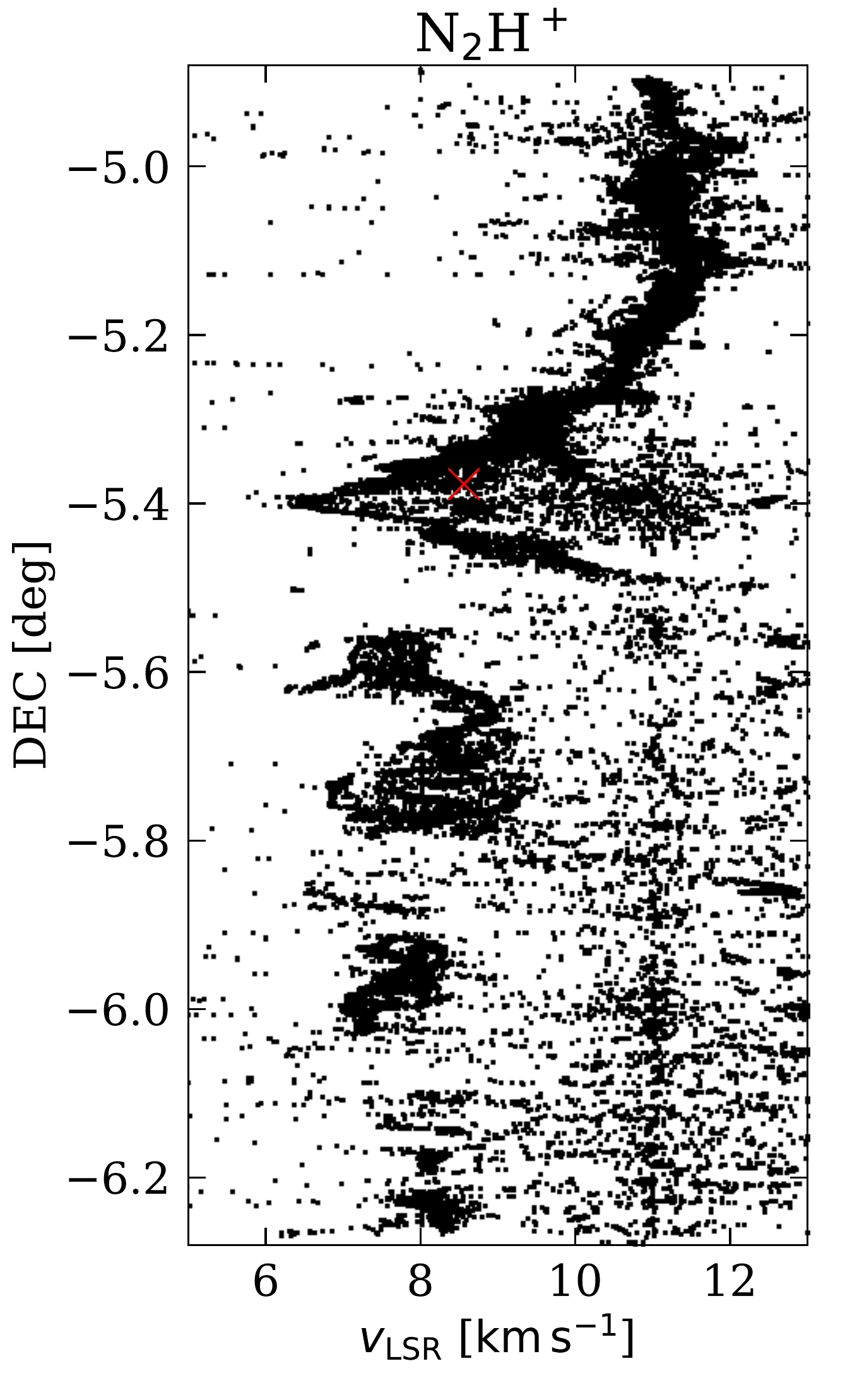}}
\caption{\textit{Left}: \nthp intensity-weighted velocity centroid as a function of $\delta$.  \textit{Right}: ``Traditional'' \nthp position-velocity diagram of the velocity centroid as a function of $\delta$, without intensity weighting. The red~{\large{$\times$}}-symbol denotes the center of mass coordinate of the ONC stars and the mean velocity of the cluster stars \citet{stutz2018a}.  The intensity-weighted diagram (left) enables us to identify structures that are either muddled or invisible in the right panel.}
\label{fig:nthp}
\end{figure*}

\begin{figure*}
  \centering
  \includegraphics[width=\textwidth]{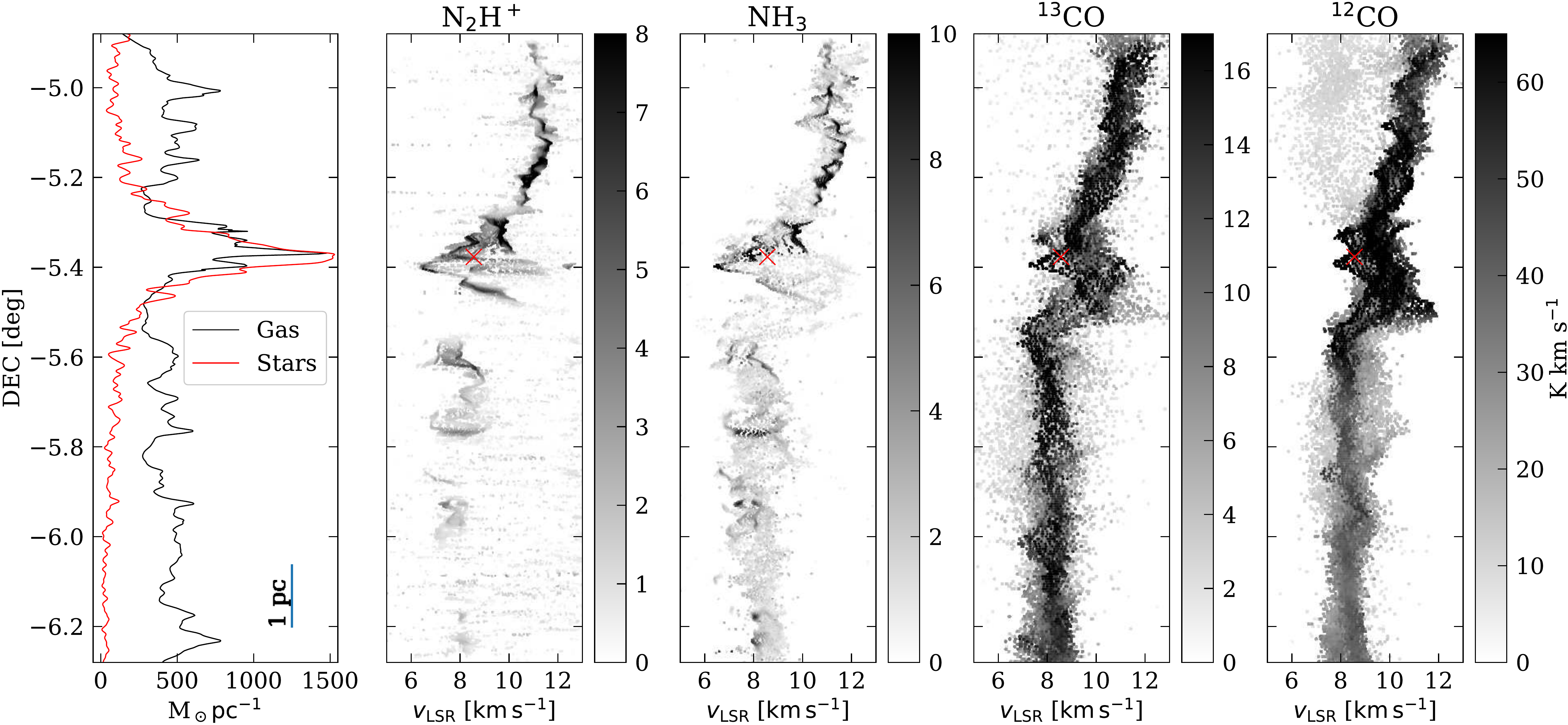} 
  \caption{From left to right: Mass per unit length as a function of $\delta$ of the gas (black curve) from \citet{stutz2018a} and young stars (red curve) from \citet{megeath2016} in the ISF; intensity-weighted velocity centroid as a function of $\delta$ for \nthp, \ammonia, $^{12}$CO and $^{13}$CO, respectively.  The position-velocity centroid (PV) diagrams are obtained from the integrated emission and velocity centroid maps for all tracers (see text). All diagrams are extracted within the \nthp observation area (see Figure~\ref{fig:n2hp_nh3}), with an R.A.\ range of $83.66^\circ\,<\alpha\,<\,84.18^\circ$ (the region indicated as the blue box in Figure~\ref{fig:co_maps}). The red~{\large{$\times$}}-symbol denotes the center of mass coordinate of the ONC stars and the mean velocity of the cluster stars \citet{stutz2018a}.  The \nthp and \ammonia PV structures are remarkably similar and clearly illustrate the small-scale velocity structure in the ISF (see text).  A blue-shifted velocity peak near the cluster position is seen in all tracers at $\delta\,\sim-5.4^{\circ}$. A hint of a double velocity component is visible in the $^{12}$CO PV diagram (see \S~\ref{sec:rotation}).}
\label{fig:pv}
\end{figure*}

\section{Intensity-weighted position-velocity diagrams}\label{sec:pv}

Our goal is to study the velocity structure along the ISF at different gas densities.  To accomplish  this, we must extract 3 principal parameters from each spectrum: integrated intensity, velocity centroid, and line width.  The line-widths will be analyzed in \S~\ref{sec:mach}.  With the first two, we construct the  intensity-weighted position-velocity (PV) shown in Figure~\ref{fig:nthp} and the second to fifth panels of Figure~\ref{fig:pv} for \nthp, \ammonia, $^{13}$CO and $^{12}$CO, respectively.  The diagrams are extracted in the area with \nthp coverage, shown as the blue box in Figure~\ref{fig:co_maps}.  All line emission outside this area is excluded from these diagrams.  

Our PV method uses the velocity integrated emission and the line velocity centroid, reducing noise in the PV diagrams and highlighting the salient velocity structure in the projected 2D space.  A comparison between our intensity-weighted method and the ``traditional'' PV diagram with no weighting shown in Figure~\ref{fig:nthp} clearly demonstrates that our method enables us to identify structures that are either muddled or invisible. Moreover, this method highlights both the small-scale and large-scale velocity structure of the ISF.  
Here we describe the specific procedure to obtain the diagrams from each tracer:
\begin{itemize}
    \item We consider the line of sight velocity centroid maps ($v_{\rm LSR}$) and integrated emission of each tracer.
    
    \item We extract the $(\alpha,\delta)$ coordinates, the $v_{\rm LSR}$ and integrated emission of each pixel in the maps that fall inside the blue box shown in Figure~\ref{fig:co_maps}.
    
    \item We plot each $v_{\rm LSR}$ point as a function of $\delta$, using the integrated emission value as grayscale. The points are plotted in order of ascending integrated emission.
\end{itemize}

It is worth noting that for points with very similar $v_{\rm LSR}$ and $\delta$, lower emission points can get eclipsed by higher emission ones.  However, this does not significantly affect the appearance of the intensity weighted PV diagram.  Below we describe in detail how we obtain the kinematic parameters for each tracer. These parameters are then used as described above to construct the intensity-weighted PV diagrams.

\subsection{\nthp}

In order to obtain the kinematic parameters of the \nthp~$(1,0)$ hyperfine line structure, we performed a fit of the \nthp~spectra using the Python module PySpecKit \citep{pyspeckit}. The module includes a built-in \nthp~algorithm. The algorithm fits all hyperfine components simultaneously and produces a model hyperfine spectrum that consists of multiple Gaussian components. The algorithm uses the  de-excitation rate coefficients from \citet{daniel2005} to determine the model spectrum considering fits of both the de-excitation rotational parameters and average opacity.  The algorithm requires an error map, a list or cube of starting values for the parameters, and a signal-to-noise ratio (S/N) threshold. We determine the errormap from emission-free channels between ${{\rm v}_{\rm LSR}=[-3.65\,\kms, -0.35\,\kms]}$.

We tested the algorithm response to different noise levels in order to choose a global S/N value. For this we performed a Monte Carlo analysis of the fitted line-width by picking a spectrum with a high ${S/N = 11}$  and adding random gaussian noise. The dispersion of the added noise distribution is given by ${\sigma=\alpha\sigma_0}$, where ${\sigma_0=0.0648\,{\rm K}}$ is the noise (rms value) of the original chosen spectrum and $\alpha$ is a number ranging from 1 to 4 in intervals of 0.05 (representing different noise levels).  We then transformed the total sigma values (${\sigma^2_{TOT} = \sigma_0^2 + \sigma^2}$) into S/N values. The process is repeated $N=1000$ times for each noise level. We found that a value of ${S/N = 4}$ produced consistent fit values. Higher values of the $S/N$ produced no change in the fitted value and its associated error. We therefore adopt the $S/N=4$ threshold throughout our \nthp hyperfine fitting procedure.  We find that the model spectra residuals are consistent with observational noise and conclude that the fitting procedure reproduces the observed spectra well.

We follow this fitting procedure to retrieve the principal kinematic parameters that we are interested in: central line velocity, linewidth, and integrated intensity, and their respective errors. We also generate excitation temperature and optical depth values, but defer analysis of these parameters to a future investigation.  We used the fitted central velocity map and integrated intensity map to generate our intensity-weighted PV diagram (similar to the procedure applied to the other tracers, see below).  Figure~\ref{fig:nthp} and the second panel of Figure \ref{fig:pv} show the resulting intensity-weighted PV diagram.

\subsection{\ammonia}

We use the \ammonia (1,1) integrated intensity map and the velocity ${\rm v}_{\rm LSR}$ map of the ISF from \citet{friesen2017}.   Panel~3 of Figure~\ref{fig:pv} shows the intensity-weighted PV diagram for \ammonia.  We highlight here the similarity between the small-scale structures in both \ammonia and \nthp; the fact that these two data-sets were acquired, analyzed, and tabulated completely independently indicates that the small-scale structures present in both are robust (see discussion below). 

\subsection{$^{12}$CO and $^{13}$CO}\label{sec:maps}

We begin by setting a 4$\sigma$ minimum S/N threshold for each spectrum. In order to do this we generate a noise map by computing the standard deviation of line-free velocity channels between ${{\rm v}_{\rm LSR}=[-10\,\kms, -2\,\kms]}$ for each spectrum. We then consider only spectra with a peak values $\geq\,4\,\sigma$. We then derive the respective velocity integrated intensity maps, radial velocity centroid maps, and linewidth maps using the Python module SpectralCube\footnote{https://github.com/radio-astro-tools/spectral-cube}.  
The intensity maps for $^{12}$CO (1-0) and $^{13}$CO (1-0) (see Figure~\ref{fig:co_maps}) are integrated over ${{\rm v}_{\rm LSR}=[0\,\kms, 16\,\kms]}$.  We use the velocity maps and intensity maps to generate intensity-weighted PV diagrams for $^{12}$CO and $^{13}$CO, shown in the fourth and fifth panel of Figure~\ref{fig:pv}. For this we extract the velocity integrated emission and line velocity centroid values at each position, and then plot the velocity centroids as a function of $\delta$, weighted by the integrated intensity, as described above for the other tracers. 

\section{Results: Velocity structure of the ISF}\label{sec:results}

Here we discuss the main features apparent in the gas velocity data, starting with the global appearance of Figure~\ref{fig:pv}.  The principal parameters that we use are the line center velocities and line widths, extracted as described above.

\subsection{Position-velocity structure as a function of density}\label{sec:velocity_struct}

Here we focus on Figure~\ref{fig:pv}.  All structures shown in this diagram were extracted within $83.66^\circ\,<\alpha\,<\,84.18^\circ$, the region indicated as the blue box in Figure~\ref{fig:co_maps}.  The first panel of Figure~\ref{fig:pv} shows the mass per unit length (M/L) distribution of the gas (black curve) and young stars (red curve) along the ISF as a function of $\delta$. The mass distribution of stars and gas in this diagram is concentrated near the ONC region \citep{stutz2016,stutz2018a}.  Due to the restricted $\alpha$ range over which we extract the M/L profiles, the gas peak near the center of the ONC is prominent (see below for comparison between the gas potential and gas velocities).  Panels 2 to 5 of Figure~\ref{fig:pv} show the radial velocity centroid of the \nthp, \ammonia, $^{13}$CO and $^{12}$CO lines as a function of $\delta$ for each tracer, sorted in order of descending critical density value (see \S~\ref{sec:data}). These four panels thus represent the gross distribution of mean line velocities along the filament, ignoring line of sight features (such as multiple velocity components). Therefore, they give us information about the plane of the sky mean radial velocity distribution. We show the center of mass and mean velocity of the ONC star cluster with a red~$\times$-symbol \citep{stutz2018a}.

The second panel of Figure~\ref{fig:pv} shows the \nthp~PV diagram of the ISF. The global structure of the dense filament is concentrated along the axis of filament with little spread in velocities at each $\delta$. The velocity structure of \nthp~is characterized by a curving velocity gradient going from ${\rm v}_{\rm LSR}\sim11\,\kms$ in the north region to ${{\rm v}_{\rm LSR}\sim7\,\kms}$ in the ONC. South of the ONC, the velocity of \nthp~is approximately constant at ${{\rm v}_{\rm LSR}\sim8\,\kms}$, with looping velocity structures which are most evident near $\delta\,\sim\,-5.7^o$.  Immediately south of the ONC, at $\delta\sim-5.5^{\circ}$ the dense filament is broken, without significant \nthp~emission.  This region presents a jump in velocity between the two ``broken'' ends of the filament of about ${{\rm v}_{\rm LSR}\sim3\,\kms}$.  This break just below the ONC was highlighted by \citet{stutz2018a} and interpreted as an indication that the instabilities that are giving rise to the general velocity structure in the ISF are propagating from North to South through the filament \citep{stutz2019}.  \citet{hacar17} show a similar analysis of the \nthp velocity structure in the ISF.   They found a velocity gradient in the proximity of the ONC with values up to $5-7\,\kms\,\pc^{-1}$, and suggest that the gas is gravitationally accelerated toward the center of the forming cluster.  The \nthp velocities near the center of the ONC (${\delta\sim-5.4^{\circ}}$) exhibit a broader spread of ${{\rm v}_{\rm LSR}\sim3\,\kms}$.  Most of the emission appears confined to the blue-shifted main peak structure with a narrow range in velocities.  As shown by the PV point for the ONC (red $\times$-symbol), the ONC star cluster has a mean velocity and postion that is offset from the dense filament gas as traced by \nthp.  The small-scale velocity fluctuations along the filament, obvious in this diagram, have a morphology analogous to wrapping or rotating structures, and persist most obviously along the entire Northern portion of the ISF.  These small scale velocity fluctuations are also observed in the southern portion of the filament (see above), although they are less prominent and have lower levels of emission.  

The third panel of Figure~\ref{fig:pv} shows the \ammonia PV diagram of the ISF.  We observe the same North to South velocity gradient seen in \nthp from ${{\rm v}_{\rm LSR}\sim11\,\kms}$ and terminating in the ONC region with ${{\rm v}_{\rm LSR}\sim7\,\kms}$. The similarities between the velocity structures, both on larger and small scales, of \ammonia and \nthp are striking considering their different critical densities (see \S~\ref{sec:data}). For \ammonia we can also observe that the filament is broken around ${\delta\sim-5.5^{\circ}}$, but whereas this region is almost completely devoid of \nthp emission in this $4^\prime$ ($\sim 0.5\,$ pc projected) interval in $\delta$, the \ammonia map traces a violent oscillation of about 6~$\kms$ toward higher ${\rm v}_{\rm LSR}$, which connects continously to the equally violent (but opposite) oscillation turning almost exactly at $\delta=-5.4^\circ$, which we already noted in the \nthp map.

The fourth panel of Figure~\ref{fig:pv} shows the $^{13}$CO PV diagram of the ISF. The distribution of velocities at each $\delta$ appear much broader than in \nthp and \ammonia, with a spread of ${{\rm v}_{\rm LSR}\sim1-2\,\kms}$ across the filament. We observe a similar North to South velocity gradient as in \nthp and \ammonia \citep[see also][ and Kong et al., 2019, submitted]{kong18} with the most blue-shifted velocity component of about ${{\rm v}_{\rm LSR}\sim8\,\kms}$ around the ONC region. Around the ONC region we observe two blue-shifted velocity peaks at ${{\rm v}_{\rm LSR}\sim8\,\kms}$ and a faint red-shifted velocity component at $\delta \sim-5.5^{\circ}$. The filament appears as one coherent structure with no signs of the break detected in \nthp and \ammonia.  The $^{13}$CO mission now bridges the gap seen at higher densities, and there is a red-shifted velocity component with the morphology of a peak near ${\delta\sim-5.5^{\circ}}$, 
which (as noted above) is both much fainter but also more finely articulated in \ammonia. 

The fifth panel of Figure~\ref{fig:pv} shows the $^{12}$CO PV diagram of the ISF. We observe a north to south velocity gradient from ${\rm v}_{\rm LSR}\sim12\,\kms$ in the north region to ${\rm v}_{\rm LSR}\sim8\,\kms$ in the ONC region.  Along the ISF, the $^{12}$CO PV structure appears to resemble a ``double helix''-type structure, echoed to some extent in the $^{13}$CO, that is most obvious in the northern portion of the filament.  Around the ONC region we see two blue-shifted velocity peaks, similar to $^{13}$CO, with velocities ${\rm v}_{\rm LSR}\sim8\,\kms$. The red-shifted velocity component seen in $^{13}$CO at $\delta\sim-5.5^{\circ}$ is now a clear structure with a red-shifted velocity peak of ${\rm v}_{\rm LSR}\sim12\,\kms$. The filament as traced by $^{12}$CO does not show any signs of the \nthp break, much like $^{13}$CO. The velocity structure of $^{12}$CO presents small scale velocity fluctuations along the filament. We observe a faint but coherent and extended blue-shifted velocity component in the northern region of the ISF at a velocity of about ${{\rm v}_{\rm LSR}\sim8\,\kms}$. This feature is apparent in the $^{12}$CO PV diagram; we carry out further analysis of it in \S~\ref{sec:rotation}.

\begin{figure}
\centering
\includegraphics[width=0.48\textwidth]{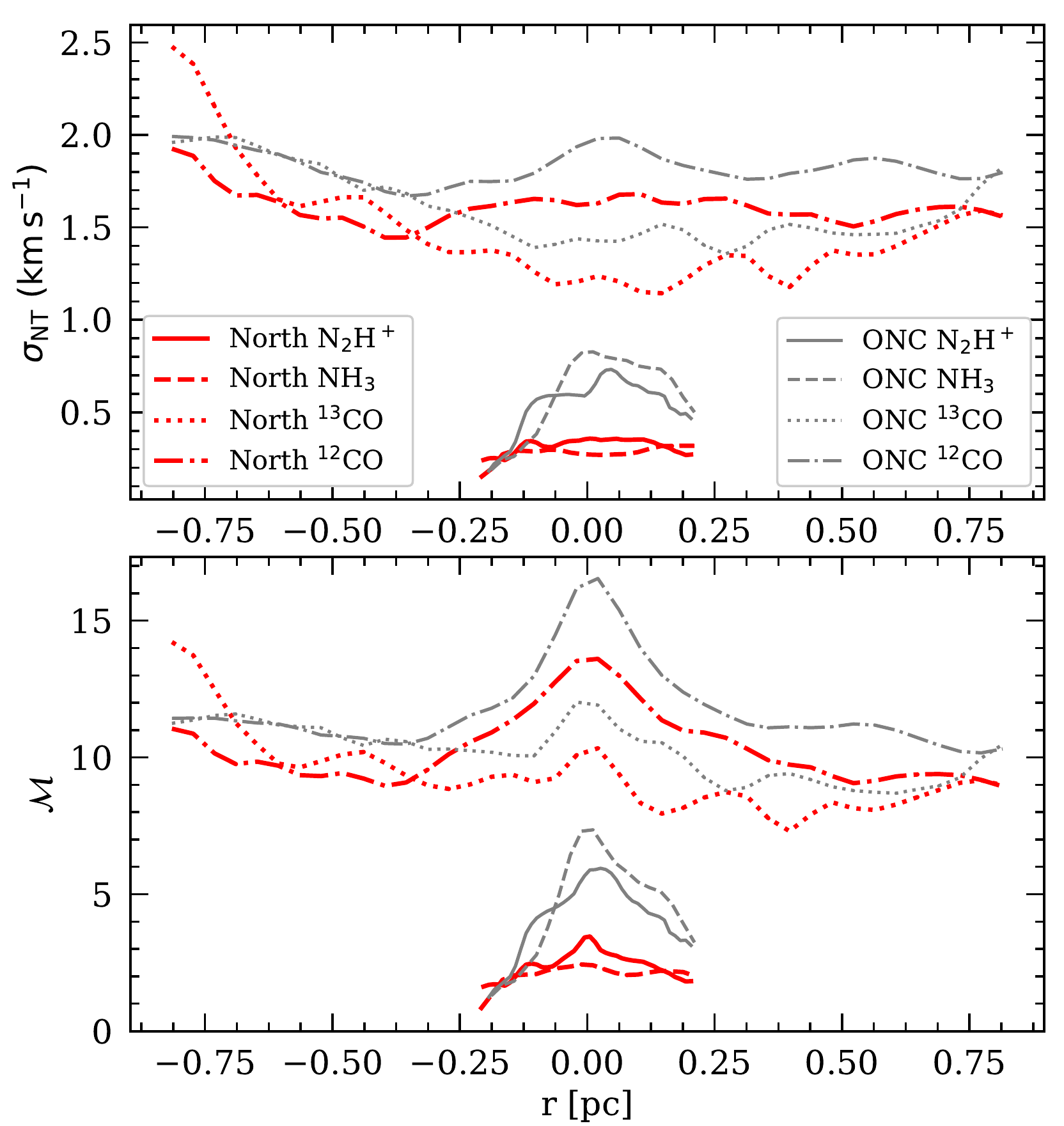}
\caption{\textit{Top:} Non-thermal component of the velocity dispersion as a function of projected radius for all tracers: $^{12}$CO is shown with the dash-dotted line, $^{13}$CO with the dotted line, \ammonia with the dashed line, and \nthp with the solid line. The red curves represent the northern region of the ISF (from $\delta=-4.9^{\circ}$ to $\delta = -5.25^{\circ}$), and the gray curves represent the OMC1 region from ($\delta =-5.25^{\circ}$ to $\delta -5.48^{\circ}$). \textit{Bottom:} Mach number as a function of projected radius for the curves shown above, assuming the temperature profile from \citet{reissl18}. Negative $r$ values correspond to the west side of the filament.}
\label{fig:radpro}
\end{figure}

\begin{figure}
\centering
\includegraphics[width=0.48\textwidth]{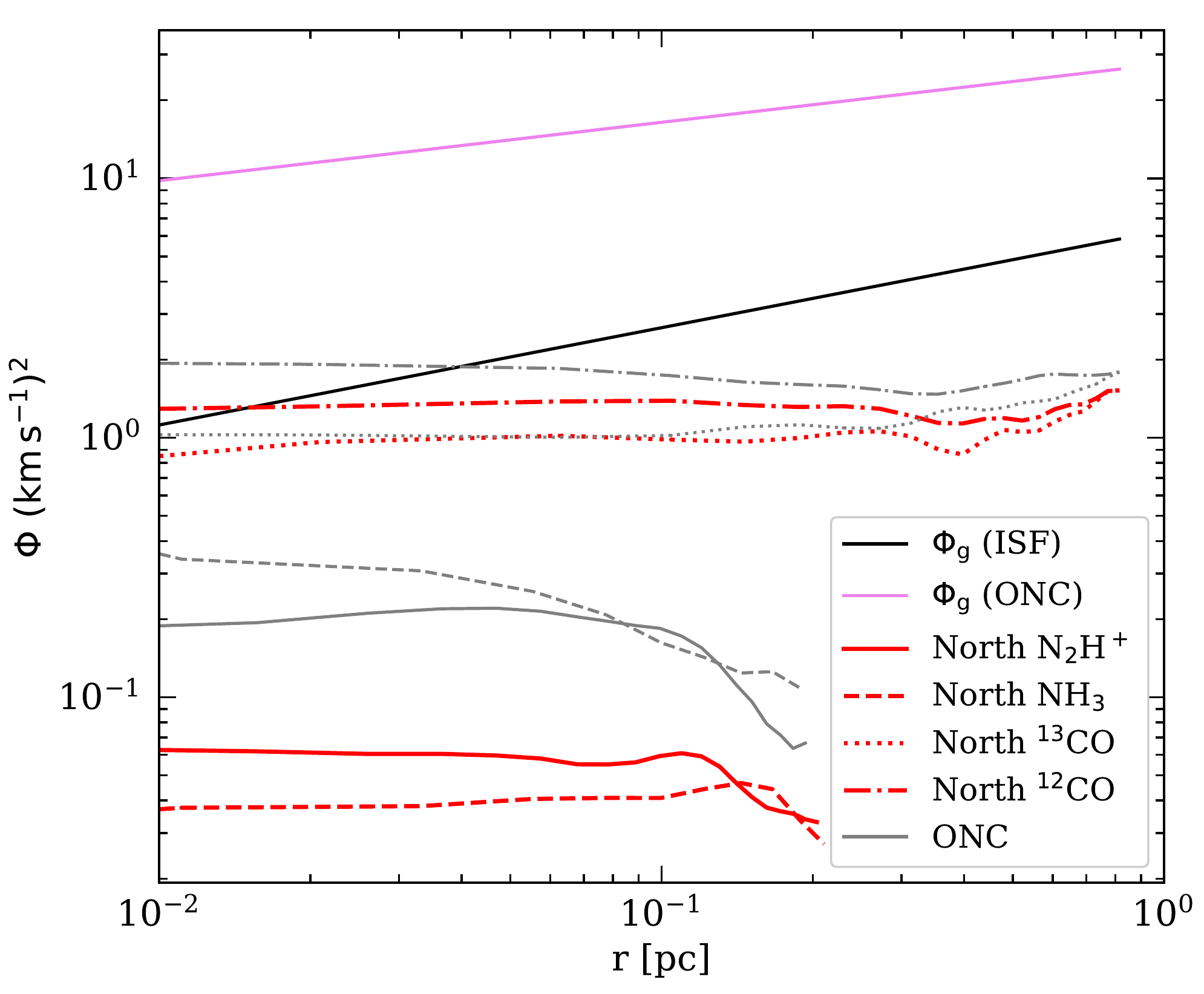}
\caption{Potential as a function of radius. The black (pink) solid curve represents the gravitational potential as a function of projected radius of the ISF (ONC) from \citet{stutz2016} \citep{stutz2018a}. The grey and red curves represent the specific kinetic energy from turbulence given by $\frac{1}{2}\sigma_{NT}^2$. These profiles are obtained by averaging both sides of the filament of the northern (red) and the OMC1 (gray) regions of the ISF.}
\label{fig:potential}
\end{figure}

We observe that all tracers show a north to south velocity gradient starting from ${\delta \sim-5.0^{\circ}}$ with velocity ${{\rm v}_{\rm LSR}\sim11\,\kms}$, terminating with a blue-shifted velocity peak of ${{\rm v}_{\rm LSR}\sim7- 8\,\kms}$ at ${\delta \sim-5.4^{\circ}}$. The peak of the stellar and total gas-mass profiles, as well as the center of mass of the ONC, are located $\sim\,0.2$~pc to north of the blue-shifted velocity peaks seen in the PV diagrams \citep[see also ][]{stutz2018a}.  The velocity gradient in the proximity of the ONC is found to be about $5-7\,\kms\,\pc^{-1}$ from \nthp~observations presented in \citet{hacar17}.  The south of the OMC1 region the velocity of the filament remains constant with ${{\rm v}_{\rm LSR}\sim8\,\kms}$ for all tracers. The overall appearance of our PV diagrams is consistent with the \citet[][]{kong18} PV diagrams extracted from higher resolution $^{13}$CO data.  Moreover, the \nthp and \ammonia data present small-scale structures that are very similar in both tracers.  These structures are also present in the the \citet[][]{kong18} diagrams, and have been previously interpreted as rotational features or possible torsional wave signatures on small scales \citep[][ and see discussion below]{stutz2019}.  

\subsection{Velocity dispersion radial profiles}\label{sec:mach}

Variations of velocity dispersion and Mach number across the ridge can give insight into cloud dynamics. Figure \ref{fig:radpro} shows the average non-thermal velocity dispersion and Mach number radial profiles for the four tracers we consider here of two regions in the ISF. The two regions were identified from the $^{12}$CO PV diagram from Figure~\ref{fig:pv}, and correspond to the northern ISF region, from $\delta=-4.9^{\circ} \rightarrow \delta = -5.25^{\circ}$, and the ONC cluster region, from $\delta =-5.25^{\circ} \rightarrow \delta =-5.48^{\circ}$. The ONC region is chosen to encapsulate the blue-shifted velocity peak seen in the four tracers (see \S~\ref{sec:velocity_struct}). The velocity dispersion radial profiles are computed from the second moment maps of each CO isotopologue and the fitted linewidth maps of 
\ammonia and \nthp accordingly. The profiles are extracted over a fixed projected radius relative to the center axis of the filament at each $\delta$, $r=0$ corresponds to the center of the filament in Figure~\ref{fig:radpro}. The center of the filament is represented using the dust ridgeline from \citet{stutz2018a}. The profiles are oriented west to east on the maps with negative $r$ values at the west side of the filament.
The \citet{stutz2018a} ridegeline shows an excellent correspondence to $^{13}$CO, NH$_3$, and in particular with the \nthp emission (see Figures~\ref{fig:co_maps} \& \ref{fig:n2hp_nh3}); therefore we adopt the \citet{stutz2018a} ridgeline as a robust tracer of the bottom of the gravitational potential well of the ISF.


To investigate the non-thermal linewidth profiles, we must subtract the thermal contribution. We compute the non-thermal velocity dispersion component as $\sigma_{NT}=\sqrt{\sigma_{obs}^2 - \frac{kT_k}{m}}$ \citep{liu2019}, where $T_k$ is the kinetic temperature  and $m$ is the mass of the species. Here we assume that the kinetic temperature $T_k$ to be equal to the dust temperature $T_{d}$ and use the dust temperature radial profile of the ISF from the radiative transfer model presented in \citet{reissl18}.  The \citet{reissl18}  temperature profile is approximated by a power law of the form ${T_{d} =9(r/{\rm pc})^{0.22}}$\,K. For each tracer the non-thermal linewidth profile is averaged in $\delta$ over each of the two regions we consider, applying a smoothing of one beam.

The velocity dispersion profiles are shown in the top panel of Figure~\ref{fig:radpro}.  The principal difference in these profiles is between different tracers (with different critical densities) rather than between regions.  The velocity dispersions of all tracers tend to be somewhat higher in the ONC region (gray curves) compared to the northern ISF region (red curves).  The average velocity dispersion in the northern ISF region (shown in red) varies from $\bar{\sigma}_{NT}=0.25$ in \ammonia to $\bar{\sigma}_{NT}=1.61$ in $^{12}$CO. This indicates that the denser gas (as traced by \ammonia and \nthp) is kinematically colder than the lower density gas ($^{12}$CO and $^{13}$CO) by a factor of $\sim\,6$.  Hence, the velocity dispersion profiles of e.g., CO alone will overestimate the dense gas line widths by factors of up to $\sim\,6$, leading to an overestimate of the non-thermal support on the dense regions where the protostars are forming \citep[e.g., ][]{stutz2016}.

The Mach number radial profiles are shown in the bottom panel of Figure~\ref{fig:radpro}.  The Mach number $\mathcal{M}=\sigma_{NT}/c_s$ is the ratio between the nonthermal component of the radial velocity dispersion and the sound speed $c_s=\sqrt{\frac{kT_k}{\mu m_H}}$, where the mean molecular weight is $\mu = 2.33$. The $\mathcal{M}$ profiles are higher in the cluster region (gray) compared to the northern region (red) of the filament. The profiles show that the Mach numbers for all tracers have a tendency to rise toward the center of the filament. The decrease in temperature toward the center of the ISF \citep{reissl18} drives the increase in ${\mathcal{M}}$ towards the center.  The variation of ${\sigma_{NT}}$ also contributes to the overall increase of $\mathcal{M}$ towards the center. The values of the Mach number show that most of the filament is highly super-sonic (${\mathcal{M}\sim5-15}$). The profiles also show that the higher density tracers (\ammonia, \nthp) are kinematically colder than the lower density gas (${^{12}{\rm CO}}$, ${^{13}{\rm CO}}$), as expected.  These profiles can be compared to simulations   provided that the simulations capture the high-mass regime of the ISF.  For a low line-mass filament example see e.g.,   \citet[][]{federrath2016}, in particular their Figure~5.

We compare the specific kinetic energy in the non-thermal linewidth profiles ($K=\frac{1}{2}\sigma_{NT}^2$) to the gravitational potential of the ISF and ONC in Figure~\ref{fig:potential}. The gravitational potential profiles of the gas in the ISF as a whole \citep{stutz2016} and in the somewhat denser ONC region \citep{stutz2018a} were derived from the \herschel column density data.  They are given by: 
\begin{equation}
{\rm ISF:}  \Phi({\rm R})=6.3(\kms)^2\left( \frac{{\rm R}}{\pc} \right)^{3/8}; 
\end{equation}
\begin{equation}
{\rm ONC:} \Phi({\rm R})= 27.6(\kms)^2\left( \frac{{\rm R}}{\pc} \right)^{0.225}.
\end{equation}
From Figure~\ref{fig:potential} it is clear that the gravitational potential profile dominates almost everywhere over the specific kinetic energy implied by the non-thermal linewidth profiles.  The only exception to this is in a small $r$ region near the center of the filament; Figure~\ref{fig:potential} shows that the specific energy in the low density gas, traced by $^{12}$CO and $^{13}$CO, is comparable to the gravitational potential, but only in the central region of the filament (with an enclosed radius of $\sim0.04\,\pc$).

It is tempting to relate these non-thermal linewidth profiles to turbulent motions.  However, because the linewidths contain both coherent (see below) and sub-beam motions, as observed in e.g., \citet{kong18,hacar18} in higher resolution data, this interpretation remains problematic.  Nevertheless, regardless of the specific interpretation of the line-width profiles, these diagrams demonstrate that the the gravitational potential dominates over the kinetic energy in the filament, showing that the filament is gravitationally bound while appearing ``supersonic''.  

\begin{figure}
\centering
\includegraphics[width=0.45\textwidth]{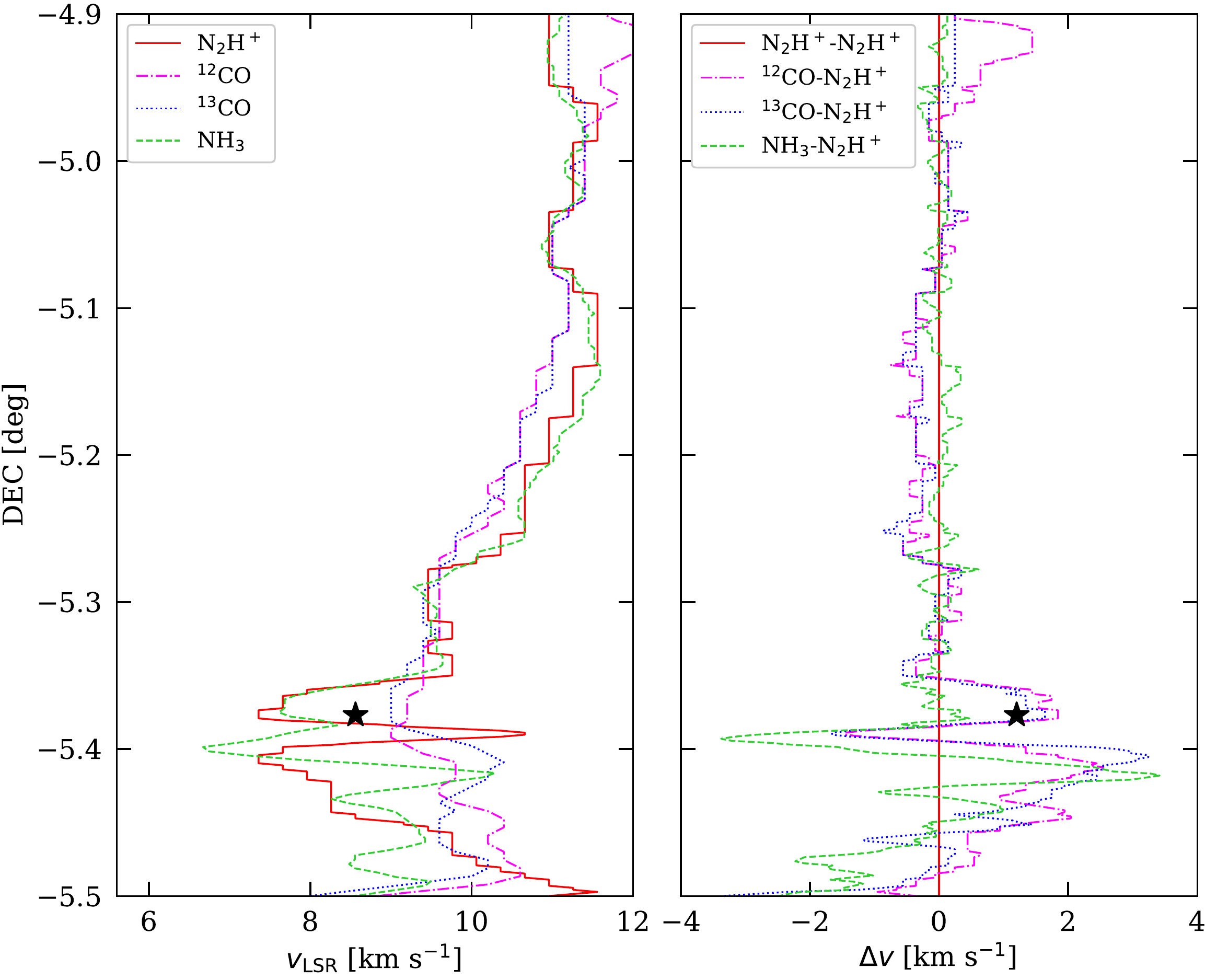}
\caption{\textit{Left}: Velocity ridgelines of the four tracers. The ridgeline is obtained from integrating the cube over the R.A. axis. Then we computed the ridgeline as the maximum intensity value at each Declination. \textit{Right}: Velocity ridgeline offsets of the four tracers with respect to the \nthp velocity ridgeline. The dot-dashed magenta curve represents $^{12}$CO, the dotted blue curve represents $^{13}$CO, the dashed green curve represents \ammonia and the red solid curve represents \nthp.  The black star indicates the location of the ONC velocity and position centroid.  }
\label{fig:ridge}
\end{figure}

\begin{figure}
\centering
\includegraphics[width=0.5\textwidth]{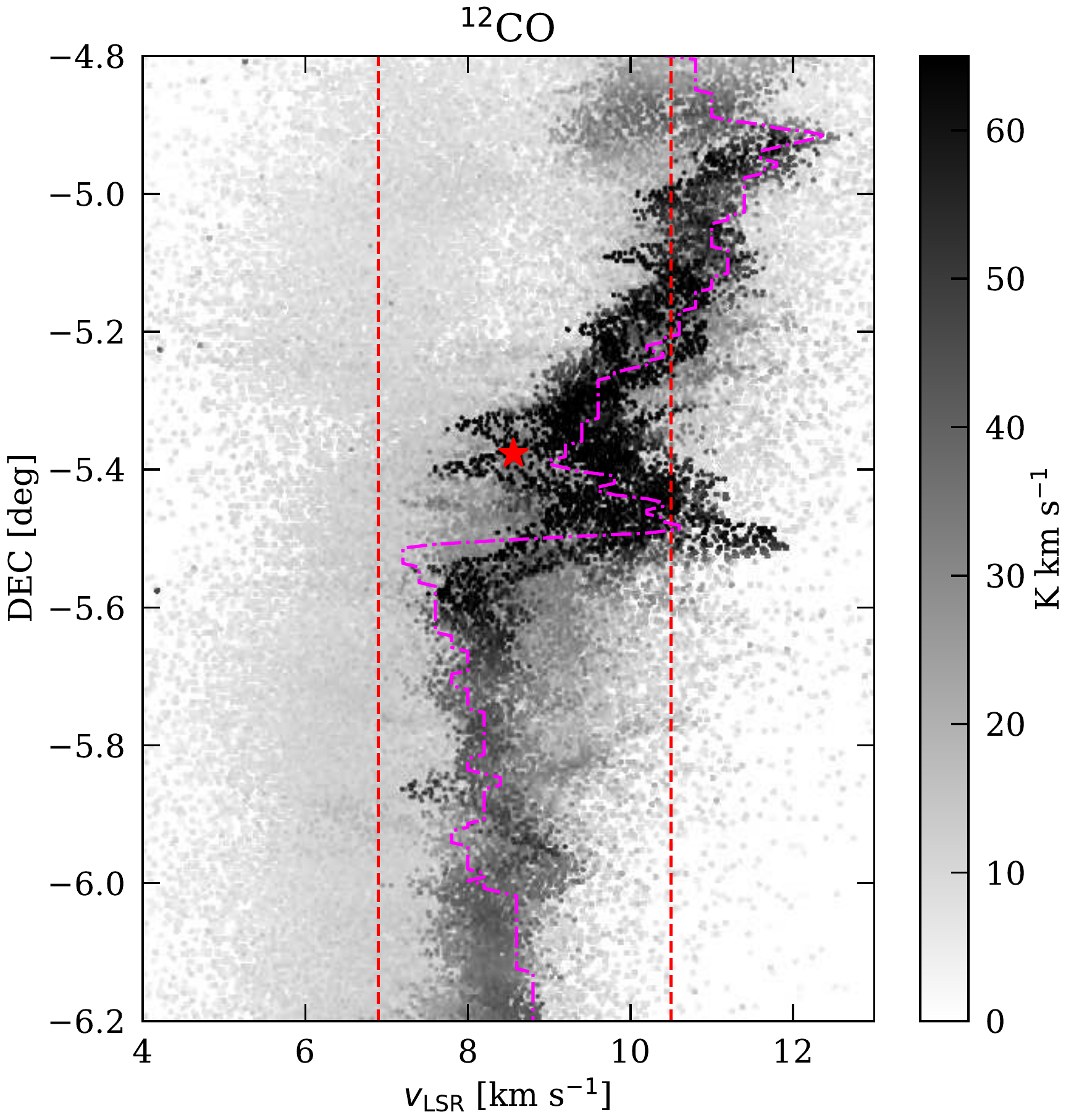}
\caption{Position-velocity centroid diagram of $^{12}$CO in the ISF. This map includes the full range of $\alpha$  values in the $^{12}$CO map. The red dashed lines indicate ${\rm v}_{\rm LSR} \sim 6.9\,\kms$ and ${\rm v}_{\rm LSR} \sim 10.5\,\kms$, the velocities of the two components in the Northern region of the ISF (seen between $\delta \sim -5.2^{\circ}$ and $-4.8^{\circ}$). The star indicates the location of the ONC velocity and position centroid, while the dashed magenta curve indicates the velocity ridgeline presented in Figure~\ref{fig:ridge} for reference.}
\label{fig:12co_pvd}
\end{figure}

\subsection{Velocity ridgelines}\label{sec:vel_ridgelines}

We aim to characterize the global kinematics of the ISF in a simple fashion. For this we investigate the velocity ridgelines of each tracer. These provide information about the global velocity structure of the ISF. Because of the comparatively low emission levels of both \ammonia and \nthp, apparent in Figure~\ref{fig:pv}, here we focus on northern ($\delta\,\geq\,-5.5\deg$) portion of the ISF where detection in all four lines are sufficiently robust to derive velocity ridgelines.  To obtain the velocity ridgeline we follow the method described in \citet{stutz2016}.  We produce PV diagrams (Figure~\ref{fig:pv}) by integrating the data cube along the $\alpha$ axis. This is accomplished by excluding areas at large impact parameters form the filament ridgeline, as described above.  At each $\delta$, we record the velocity corresponding the maximum intensity.  To suppress noise we apply smoothing of two beamsizes to the ridgeline for each tracer.  

Figure \ref{fig:ridge} shows the velocity ridgelines of the four tracers (left panel) and the difference relative to the \nthp~velocity ridgeline (right panel).  This figure shows that the four tracers have very similar ridgelines along the filament, except in the ONC. The velocity ridgelines of all tracers present a north to south velocity gradient from $\sim\,11\,\kms$ to $\sim9-7\,\kms$, similar to the wave-like pattern shown in \citet{stutz2019}. In the ONC the velocity ridgelines are more blue-shifted (${{\rm v}_{\rm LSR}\sim8\,\kms}$) than in the north, also seen in Fig.~\ref{fig:pv}. Moreover, as also seen in  Fig.~\ref{fig:pv}, \ammonia follows \nthp closely along the northern portion of the filament.  In the center of the ONC (near $\delta\,\sim\,5.4$\degree), we observe larger differences between the ridgelines, which we interpret as velocity differences as a function of density in the more extreme central environment of the cluster.  

These velocity gradients and morphology are very similar to those presented in \citet{kong18}, see e.g., their Figures~20-22.  These authors observe the cloud at higher resolution ($8\arcsec$) using $^{12}$CO, $^{13}$CO, and C$^{18}$O (1-0). Despite the differences in both resolution and tracers, the overall morphology is similar to our ridgelines (see also Figure~1 of \citealt{hacar17}).  Moreover, in a similar analysis, \citet{wu2018} showed similar fluctuations and offsets between $^{12}$CO (2-1) and \nthp in their velocity ridgelines of the ISF. They propose that these fluctuations may be rooted in the oscillation of the filament from the slingshot mechanism \citep{stutz2016,stutz2019}. In contrast to the \citet{wu2018} analysis, here we use the \nthp as our velocity reference in the right panel of Figure~\ref{fig:ridge}, as the \nthp is tracing higher densities and thus is more likely to provide the velocity relative to the bottom of the filament potential (as apposed to $^{12}$CO; \citealt{wu2018}).  Nevertheless, the magnitude of the  $^{12}$CO and \nthp velocity fluctuations are consistent along the ISF (${\Delta{\rm v}\sim1-2\,\kms}$) in both works. \citet{wu2018} also found a $0.7\,\kms\,\pc^{-1} (= 0.7\,{\rm Myr}^{-1})$ velocity gradient along the filament in the center of ONC which they propose may arise from the global contraction of the ISF, also discussed in \citet{hacar17}.  

\subsection{Rotation of the filament}\label{sec:rotation}

The $^{12}$CO panel of Figure~\ref{fig:pv} shows a double velocity component in the northern region of the ISF.  The morphology is remiscent of a ``tuning fork'', i.e., two vertical loci in velocity that merge into a sigle common velocity near the center of the ONC.  We further investigate this signature generating a PV diagram that covers the full range in $\alpha$ of the $^{12}$CO emission of the ISF, shown in Figure~\ref{fig:12co_pvd}.  This figure shows that in the Northern portion of the ISF, the two main velocity components in the $^{12}$CO gas are located at about 
${\rm v}_{\rm LSR} \sim 6.9\,\kms$ and ${\rm v}_{\rm LSR} \sim 10.5\,\kms$, respectively, indicated with the dashed red lines.  In both figures, the blueshifted component is significantly fainter than the redshifted one, and the bright component merges with the faint component at or just above the ONC. The faint blueshifted component persists below the ONC and into the L1641 cloud, where the velocity gradient disappears on all tracers. As described above, in Figure~\ref{fig:pv} we restrict the $\alpha$ range of integration; meanwhile, in Figure~\ref{fig:12co_pvd} we include the full range in $\alpha$ presented in Figure~\ref{fig:co_maps}.  Comparison between these two figures, and in particular the persistence of the feature in both indicates that this velocity split in the Northern portion of the ISF is persistent over a large range of distances from the spine of the filament.  

This is, to the best of our knowledge, the first place that this velocity signature and corresponding morphology have been presented.  Our PV diagrams show the radial velocity stucture of the ISF in the plane of the sky.  The spatial separation provides a rough estimate of the spatial separation of the two components of ${\rm r}=1.3\,\pc$.  Assuming these two velocity components arise from gas rotation, we apply a simple circular velocity model to interpret this signature, and compute the angular velocity as: 
\begin{equation}
\omega=\frac{\rm \delta v/2}{\rm r} = 1.4\,\kms{\rm pc}^{-1} = 1.4\,{\rm Myr}^{-1}. 
\end{equation}
The presence of rotation in the ISF has been previously noted by \citet{tatematsu1993} based on $^{13}$CO observations, where they found a velocity gradient of $2-4\,\kms\,\pc^{-1}$ across the major axis of the filament. \citet{hanawa1993} inferred an angular velocity of $\Omega \sim 1-1.5\times10^{-13}\,{\rm s}^{-1}$ that in combination with the proposed helical magnetic field of the ISF \citep{heiles97} may produce significant effects on the filament \citep{hanawa1993}; see also \citet{dom18}. Alternatively, the origin of the two velocity components in the ISF may be unrelated to rotation, and instead rooted in other causes, such as cloud-envelope velocity diferences, infall, or cloud-cloud collisions \citep[e.g., ][]{fukui2018}.  In this later scenario  \citet{fukui2018} propose that about 0.1~Myr ago a lower mass cloud collided with the main cloud, triggering massive star formation.  Meanwhile, their $^{12}CO$ position-velocity diagram (their Figure~8) does not show the fainter blueshifted northern velocity component presented here.   However, their theoretical simulations, albeit with approximately spherical initial conditions, do apear to produce in some cases a double velocity distribution.  It remains to be investigated if in the framework of cloud collisions the velocity features presented here can be reproduced while accounting for the various robust constraints, principally the filamentary geometry, the gravitational potential, and corresponding timescales of star formation. 

We observe several blue-shifted velocity peaks in the PV diagram of Figure~\ref{fig:12co_pvd}. We analyze these in detail in Appendix~\ref{sec:peaks_app}.


\section{Summary}\label{sec:sum}

In this paper we have used public observations of different molecular tracers, namely $^{12}$CO~($1-0$), $^{13}$CO~($1-0$), \ammonia~($1,1$) and \nthp~($1-0$), along with other observations, in order to characterize the velocity structure of the gas associated with the ISF and processes of star cluster formation.  Here we summarize our main results: 

\begin{itemize}
    \item We present intensity-weighted position-velocity diagrams in Figures~\ref{fig:nthp} and~\ref{fig:pv} of the ISF. Our method reveals large and smaller scale structures in the gas that where previously difficult to identify.
    \item On larger scales we observe a north to south velocity gradient that terminates with a blue-shifted velocity peak near the center of the ONC that is detected over the large range in critical densities of our tracers, consistent with the previously described wave-like behaviour of the ISF \citep{stutz2016, stutz2018a, stutz2019, kong18}.
    \item On smaller scales we detect twisting and turning structures present in \ammonia and \nthp, these structures have short timescales and give the impresion of a torsional wave.
    \item We present analysis of the linewidths of all tracers in Figure~\ref{fig:radpro} and~\ref{fig:potential}.  We find that the linewidths are strongly dependent on the tracer and supersonic, with lower density tracers having larger line widths, Mach numbers, and specific kinetic energy than higher density tracers over the radii where the tracers can be compared. 
    \item The inferred specific kinetic energy from the linewidths are much lower than the gravitational potential in both the ONC region and the northern part of the ISF \citep{stutz2016,stutz2018a}, suggesting that the cloud is gravitationally bound.  The fact that the gravitational potential dominates over the gas motions implies that either the cloud is collapsing under its own gravity or that additional support, such as that provided by magnetic fields and/or rotation, are required to maintain plausible efficiencies.
      \item We report the presence of two velocity components seen in the $^{12}$CO PV diagram in the northern region of the ISF, if we interpret these components as circular rotation we find an angular velocity of $\sim1.4\,\myr$.

\end{itemize}

The analysis of the data presented here suggests a global view of massive filament and cluster formation that is more complex than previously appreciated.  We detect various features, both on large and small scales, that cry out for an explanation.  Specifically, the northern portion of the ISF has obviously distinct properties compared to the southern half.  We suggest that, as previously proposed, the action of the magnetic field is of  primary importance, along with gravity, in shaping the present-day properties of the ISF.

\section*{Acknowledgements}

VGL and AMS thank the anonymous referee for comments that helped improve the clarity of this work, and HongLi Liu for helpful suggestions. VGL gratefully acknowledges support from CONICYT Beca Magister Nacional 22182160, the Chilean BASAL CATA grant PFB-06/2007 and Dominik Schleicher.  VGL also thanks IC, FP, LK and GT for their support and contributions to this work.  AS acknowledges funding through Fondecyt regular (project code 1180350), ''Concurso Proyectos Internacionales de Investigaci\'on'' (project code PII20150171), and Chilean Centro de Excelencia en Astrof\'isica y Tecnolog\'ias Afines (CATA) BASAL grant AFB-170002. This research has made use of the SIMBAD database, operated at CDS, Strasbourg, France.

\bibliographystyle{mnras}
\bibliography{ref,ref2}

\appendix

\section{Velocity peaks in $^{12}$CO versus YSO and protostar positions}\label{sec:peaks_app}

\begin{figure*}
\centering
\subfloat{\centering\includegraphics[width=0.48\textwidth]{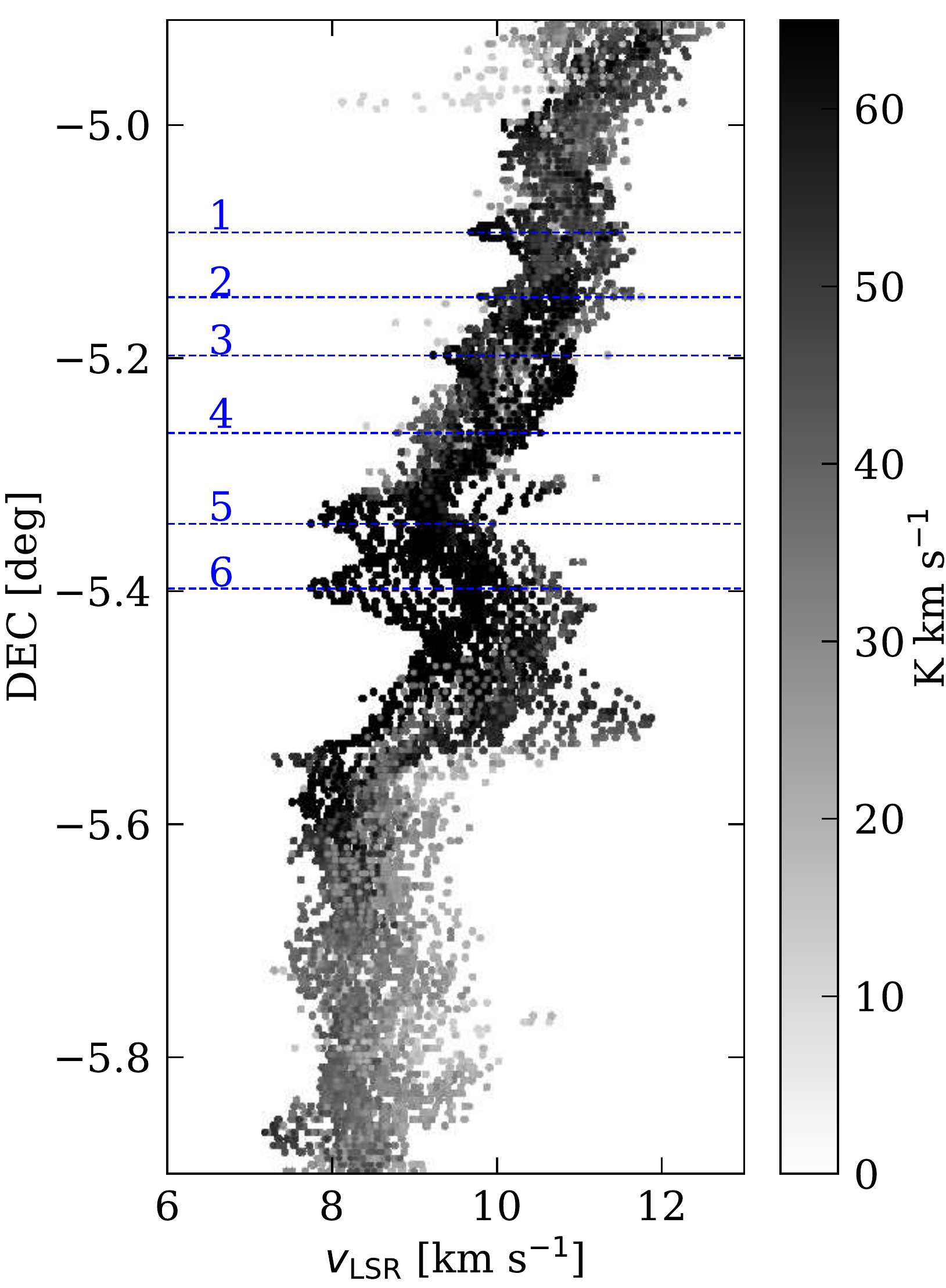}}%
\subfloat{\centering\includegraphics[width=0.5\textwidth]{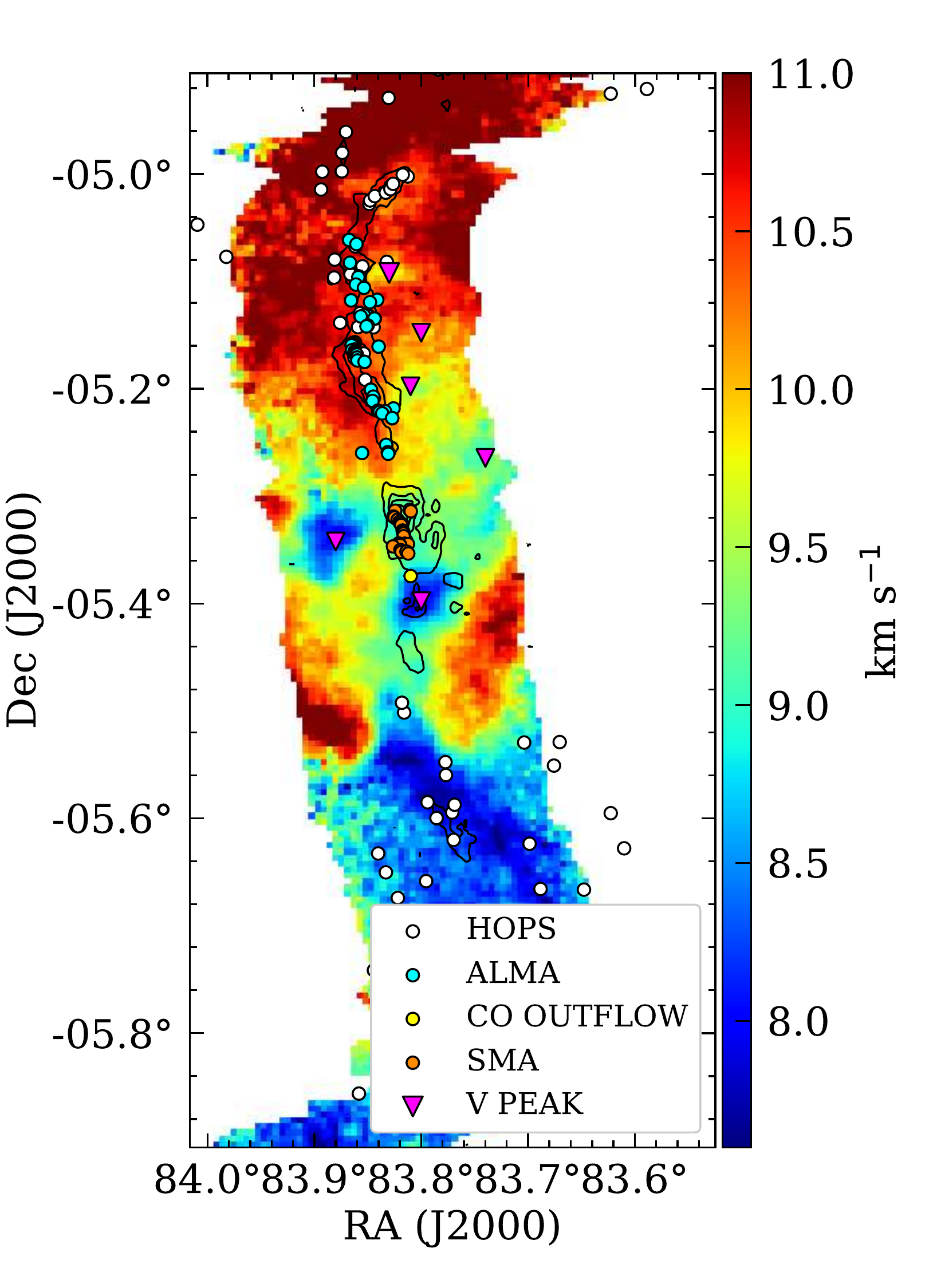}}
\caption{{\it Left}: $^{12}$CO PV diagram of the area following the dust ridgeline from \citet{stutz2018a} at a fixed projected radius of ${\sim0.8\,\pc}$. {\it Right}: Moment~1 map of the $^{12}$CO cube showing the area that follows the dust ridgeline, where there is significant \nthp emission. The $\alpha$ and $\delta$ position of the velocity peaks are shown with large magenta triangles.  The  \nthp emission is indicated with black contours.  The circles are protostar and point sources from different catalogs in the ISF region: white~$\to$~HOPS catalogue \citep{furlan2016}, cyan~$\to$~ALMA \citep{kainulainen2017}, yellow~$\to$~the Orion KL region \citep{zapata2009}, and orange~$\to$~protostellar cores observed with the SMA \citep{teixeira2016}. The velocity peaks are shown with the magenta triangles.}
\label{fig:cat}
\end{figure*}

Here we present detailed analysis of the velocity peaks first identified from Figure~\ref{fig:12co_pvd}. The left panel of Figure~\ref{fig:cat} shows the PV diagram of $^{12}$CO extracted by following an area within ${{\rm r}=0.8\,\pc}$ from the dust ridgeline from \citet{stutz2018a}, in order to track the dense filament and the well of the gravitational potential.  We identify six blueshifted velocity peaks from the PV diagram, indicated in Figure~\ref{fig:cat}. Each peak is numbered and the corresponding Declination is indicated with horizontal lines. The peaks appear to be spaced regularly, with approximate mean spatial separations of $\delta\,\sim 3.6^{\prime}$ or $\sim 0.44\,\pc$.  To investigate the possible origin of these velocity peaks, we identify their $\alpha$, indicated in the right panel of Figure~\ref{fig:cat} with triangles.  These velocity peaks are located outside of the regions with protostars and dense cores, which follow densest regions traced by the dust ridgeline. Specifically, we compare protostar positions using the HOPS catalog from \citep{stutz13,furlan2016}, protostellar cores observed with ALMA at $3\,{\rm mm}$ continuum of the northern part of the ISF from \cite{kainulainen2017}, protostellar cores identified in the OMC1 North region observed with the Submillimeter Array (SMA) at ${1.3\,{\rm mm}}$ from \cite{teixeira2016}, and the outflow source in the Orion KL region from \cite{zapata2009} observed with the SMA (see Figure~\ref{fig:cat}, right panel).  None of the peaks except Peak~6, near the center of ONC and the BN/KL explosion, appear to be coincident with the overlayed protostars and dense cores.  

As describe above, we observe nearly periodic peaks in the $^{12}$CO PV diagram.  One possible explanation for these is fortuitously periodic outflows from Young Stellar Objects (YSOs).  To identify the driving source (or lack thereof) of the regularly spaced velocity peaks, we crossmatch the velocity peaks using the SIMBAD Astronomical Database \citep{wenger2000}, the HOPS catalog from \citet{furlan2016}, the YSO catalog from \citet{megeath2012,megeath2016}, and the HOPS sources associated with CO outflows in the OMC2/3 region \citep{kainulainen2017} to investigate possible outflow sources in the vicinity of each peak. In the appendix we present detailed discussion the source crossmatch search for each velocity peak. In summary,
we find that Peaks 2, 4 and 5 are not associated with any cataloged YSO outflow.  Peaks 1 and 3 cannot be ruled out as being associated with YSO outflow activity in their vicinity, while Peak 6 is associated with the ONC. Below is a more detailed analysis of each velocity peak.

\begin{itemize}

\item \noindent{\bf Peak 1}: Located at $\alpha$~=~05h35m19.2s, $\delta=-05^{\circ}05^\prime24^{\prime\prime}$. SIMBAD inspection reveals that the closest source is an x-ray source identified with Chandra (source I151 from \citealt{tsujimoto2002}) at a separation of 12.5$^{\prime\prime}$ from Peak~1, at $\alpha$=~05h35m19.75s, $\delta$~=~$-$05\degree05$^\prime$33.4$^{\prime\prime}$). The second closest source is a dense core located at $\alpha$~=~05h35m20.2s, $\delta$~=~$-$05\degree05$^\prime$14$^{\prime\prime}$ identified in H$^{13}$CO$^{+}$ emission \citep{ikeda2007} at a separation of $17.9\arcsec$ from Peak~1. These two sources do not appear  to be associated with an outflow.  

We identify a protostar at a separation of about $\sim30^{\prime\prime}$ from the peak in the HOPS catalog \citep{furlan2016}. The protostar is named HOPS082 and is located at $\alpha$~=~05h35m19.73s, $\delta$~=~$-$05\degree04$^{\prime}$54.6$^{\prime\prime}$. The HOPS082 protostar is classified as a flat spectrum source, indicating that is going through the process of evolving towards a T-Tauri like disk star and is more evolved than its younger protostellar siblings. HOPS082 has a bolometric temperature of T$_{\rm BOL}$=116.4~K, a bolometric luminosity of L$_{\rm BOL}$=2.4~\msun and an inner envelope mass of M$_{\rm env}$=5.5$\times10^{-2}$~\msun \citep{furlan2016}. We find no indication in the IRAC images from \citet{megeath2012,megeath2016} of extended emission associated with possible outflow activity in this source. Given the low luminosity, low envelope mass, the evolved state and no outflow indication we discard HOPS082 as a source of the velocity feature observed in Peak~1. 

On the other hand, further inspection of the SIMBAD sources around a radius of 90$^{\prime\prime}$ centered on Peak 1 reveals a series of H$_2$ flows identified in \citet{yu1997} that are associated with the dust condensations MMS9 ($\alpha$~=~05h32m58.2s, $\delta$~=~$-$05\degree07$^{\prime}$35$^{\prime\prime}$) and MMS10 ($\alpha$~=~05h33m04.5s, $\delta$~=~$-$05\degree07$^{\prime}$34$^{\prime\prime}$) from \citet{chini1997}. MMS9 presents a collimated east to west chain of H$_2$ knots that terminates 3$^{\prime}$ west of MMS9. The $^{12}$CO velocity map of Orion A from \citet{kong18} shows in more detail the region near peak 1 and reveals an outflow structure at ${\delta\sim-5^{\circ}06'}$ that extends in the west to east direction. The presence of such outflows in the vicinity of Peak 1 indicate that this velocity feature could be caused by molecular outflows originating from the \citet{chini1997} sources.

\item \noindent {\bf Peak 2}: Located at $\alpha$~=~05h35m09.6s, $\delta$~=~$-$05\degree08$^{\prime}$24$^{\prime\prime}$. Inspection on the SIMBAD database shows that the closest source is a YSO located at $\alpha$~=~05h35m10.7s, $\delta$~=~$-$05\degree08$^{\prime}$25.8$^{\prime\prime}$, at a distance of 16.5$^{\prime\prime}$ from peak 2. The YSO was identified with SCUBA-2 and has a mass of 0.23~\msun and a temperature of 10.4~K \citep{salji2015}. There is no evidence for outflow activity.  The second closest source is a pre-main sequence star located at $\alpha$~=~05h35m10.72s, $\delta$~=~$-$05\degree08$^{\prime}$16.95$^{\prime\prime}$, at a distance of 18.2$\arcsec$ to peak 2 identified in infrared by \citet{jones1994}. This source is detected in the Gaia catalog, indicating that it is not subject to elevated levels of extinction due to a large envelope-outflow system. This source is detected in x-rays in the \citet{getman2017} catalog.  This source is also identified in the catalog of \citet{megeath2012,megeath2016} and has a steeply falling SED through the IRAC bands and exhibits no morphological indications of outflow activity in the IRAC images.  Two more YSOs from \citet{megeath2012,megeath2016} appear within a radius of 90$^{\prime\prime}$ from the position of Peak~2. These share similar characteristics as the YSO discussed above, that is, steeply falling SEDs and no morphological indications of outflow activity in the IRAC images. Extensive Vizier and SIMBAD searches do not reveal any outflow near the location of Peak~2.

\item \noindent{\bf Peak 3}: Located at $\alpha$~=~05h35m12s, $\delta$~=~$-$05\degree11$^{\prime}$24$^{\prime\prime}$. The closest source is 27.4$^{\prime\prime}$ from Peak~3, an x-ray source (I76) located at $\alpha$~=~05h35m10.32s , $\delta$~=~$-$05\degree11$^{\prime}$12.9$^{\prime\prime}$.  I76 was identified in x-ray using Chandra \citep{tsujimoto2002}, but does not have enough information for classification. The source does not present any obvious indication of outflow activity.  The second closest source is a low mass star located at $\alpha$~=~05h35m12.7s,  $\delta$~=~$-$05\degree12$^{\prime}$00.68$^{\prime\prime}$, at a distance of 38.16$^{\prime\prime}$. This source was first identified in NIR by \citet{jones1994}. The star has a mass of about 0.1~\msun \citep{dario2012}. Given its evolutionary state, low mass, and no apparent association to an outflow we do not regard this as a probable source of velocity Peak~3.  One YSO from \citet{megeath2012,megeath2016} is located at a distance of about 70$^{\prime\prime}$ from the Peak~3 position ($\alpha$~=~05h35m07.52s, $\delta=-05^{\circ}11'14.14''$) and is classified as a pre-main sequence star with disk. This YSO shows no indication of outflow activity in the IRAC bands. Further SIMBAD inspection shows that this source is classified as a rapid irregular variable \citep{samus2017}. The source is detected  in the Gaia DR2 catalog. It is also identified with Chandra, is variable, and was classified as a Class~II protostar \citep{tsujimoto2002}. Given the evolutionary state of this YSO and the lack of evident association to an outflow we discard this object as a source of Peak~3.

We find three YSOs from the HOPS catalog that have been associated with CO outflows near this region \citep{kainulainen2017,furlan2016}. These correspond to HOPS60 ($\alpha$~=~05h35m23.3s, $\delta$~=~$-$05\degree12$^{\prime}$03.15$^{\prime\prime}$) classified as a Class~0 protostar with a bolometric luminosity of 21.93~\lsun, a bolometric temperature of 71.5~K and an envelope mass of 0.135~\msun. This YSO is associated with the FIR6b outflow from \citet{shimajiri2009} at $\alpha$~=~5h35m21.37s, $\delta$~=~-5\degree12$^{\prime}$05$^{\prime\prime}$. The outflow orientation is East-West, and inspection of the position and extent reveals a very similar signature to the one present in our $^{12}$CO velocity map, allowing us to conclude that Peak~3 is likely produced by the outflow associated with the FIR6 source.  

\item \noindent{\bf Peak 4}: Located at $\alpha$~=~05h34m57.6s, $\delta=-05^{\circ}15^{\prime}36^{\prime\prime}$. The closest source is an infrared source located at $\alpha$~=~05h34m58.83s, $\delta$~=~$-$05\degree15$^{\prime}$35.7$^{\prime\prime}$, at a distance of 18.37$^{\prime\prime}$ from the position of Peak~4 \citep{tsujimoto2003}. We find no indication of outflow activity tied to this source. The second closest source is a Star located at $\alpha$~=~05h34m56.37s, $\delta$~=~$-$05\degree15$^{\prime}$28.4$^{\prime\prime}$, at a distance of 19.9$^{\prime\prime}$ from the position of Peak~4. The star was identified by \citet{tsujimoto2003} and is also observed by 2MASS \citep{cutri2003}. We find no evidence of outflow activity associated to this source.  We find two YSOs from \citet{megeath2012, megeath2016} located at $\sim\,45\arcsec$ from Peak~4. The coordinates are $\alpha$~=~05h34m56.67s,  $\delta$~=~$-$05\degree14$^{\prime}$52.26$^{\prime\prime}$, and $\alpha$.~=~05h35m00.22s, $\delta$~=~$-$05\degree15$^{\prime}$58.97$^{\prime\prime}$ respectively. The first YSO (also identified in Gaia DR2) classified as a pre-main sequence star with disk \citep{megeath2012} and is a periodic variable with a rotation period of 2.65~d \citep{rodriguez2009}. The second YSO is classified as a pre-main sequence star with disk \citep{megeath2012} and presents variability due to rotation with a period of 1.25~d \citep{rodriguez2009}. Neither of these sources show any morphological indication of outflow activity in the IRAC images.

\item \noindent{\bf Peak 5}: Located at $\alpha$~=~05h35m28.8s, $\delta$~=~$-$05\degree20$^{\prime}$24$^{\prime\prime}$. SIMBAD inspection of the vicinity of the peak shows that the closest source is found at a distance of 3.7$\arcsec$, located at $\alpha$~=~05h35m28.8s,  $\delta$~=~$-$05\degree20$^{\prime}$20.3$^{\prime\prime}$ \citep{muench2002}. This star shows no evidence for outflow activity.  We find two YSOs from \citet{megeath2012,megeath2016} located at about 45\arcsec from Peak 5. The first YSO (at $\alpha$~=~05h35m26.17s,  $\delta$~=~$-$05\degree20$^{\prime}$06.04$^{\prime\prime}$) is identified in Gaia DR2 and 2MASS. The YSO presents variability with a rotation period of 8.55~d \citep{rodriguez2009} and is classified as a pre-main sequence star with disk \citep{megeath2012}. The second YSO (at $\alpha$=~05h35m31.39s, $\delta$~=~$-$05\degree20$^{\prime}$17.05$^{\prime\prime}$) is classified as a pre-main sequence star with disk \citep{megeath2012}. Neither of these protostars show any morphological indication of outflow activities in the IRAC images.  Given the more evolved state of the YSOs close to Peak~5 and the lack of evicence for outflow activity in the vicinity, we conclude that there is no indentified outflow source associated to Peak~5.  In contrast to Peak~6 (see below), Peak 5 is most prominent in $^{12}$CO, less so in $^{13}$CO, and appears coincident with a dense knot in \ammonia nad \nthp, located just above but near the position-velocity of the ONC (see red $\times$-symbol in Figure~\ref{fig:pv}).

\item\noindent{\bf Peak 6}: Located at $\alpha$~=~05h35m12s, ${\delta=-05^{\circ}23^{\prime}24^{\prime\prime}}$. SIMBAD inspection of the vicinity of the peak reveals a very crowded region with sources related to the ONC, as expected.  This peak lies near the center of the cluster, in the Orion BN/KL region \citep{becklin1967,kleinmann1967}. This region is characterized by an explosion caused by the close interaction between two or more protostars about 500\,\yr ago \citep{zapata2009,bally2017}. This explosion induced a wide outflow with a scale of $\sim1'$ that is spherically symmetric distributed around the explosion center \citep{bally2017}. The center of the explosion is located at $\alpha$~=~05h35m14.37s, $\delta$~=~$-$05\degree22$^{\prime}$27.9$^{\prime\prime}$ \citep{zapata2009} at a separation of $\sim\,79.2 ^{\prime\prime}$ from Peak 6. Taken at face value, the position of Peak~6 is inconsistent with the explosion from the Orion BN/KL explosion \citet{bally2017}.  However, the low velocity component of the outflow (within ${10\,\kms}$ from the OMC1 rest velocity ${{\rm v}_{\rm LSR}\sim9\,\kms}$) appears elongated from southeast to northwest, with a $\sim\,1^\prime$ long wide blueshifted bubble that extends southeast with velocity ${{\rm v}_{\rm LSR}\sim7\,\kms}$. This suggests that Peak~6 may in fact be related to the BN/KL explosion. However, and in contrast to Peak~5 discussed above, the detection of Peak~6 in all tracers (see Figure~\ref{fig:pv}), from the lower to high density gas, indicates that that interpretation of Peak 6 as driven by the BN/KL explosion may be problematic.  The positon-velocity structure that we observe for Peak~6 is consistent across all tracers, which indicates that larger-scale and more global filament dynamics are responsible for shaping the observed morphology.  

\end{itemize}

Another possibility for the origin of the peaks is periodic infall, while another is that the peaks are related to the wave phenomenon in the ISF.  The characteristic distance between the peaks is $\sim\,0.44\,\pc$.  Meanwhile the characteristic velocity relative to the main local filament velocity is $\sim\,0.5\kms$.  Together, these provide an estimate of the characteristic timescales of $\sim\,1$~Myr.  This is very similar to the other timescale estimates above (see \S~\ref{sec:rotation}), and indicates that both the short- and long-range phenomena that are associated with similar timescales and may have a causal connection, as would be expected from wave-like behavior.


\bsp
\label{lastpage}
\end{document}